\documentclass[prd,eqsecnum,twocolumn,amsfonts,showpacs]{revtex4}

\usepackage{graphicx}

\usepackage{bm}

\newcommand{\beq}{\begin{equation}}
\newcommand{\eeq}{\end{equation}}
\newcommand{\beqs}{\begin{eqnarray}}
\newcommand{\eeqs}{\end{eqnarray}}
\newcommand{\lsim}{\mathrel{\raisebox{-
.6ex}{$\stackrel{\textstyle<}{\sim}$}}}
\newcommand{\gsim}{\mathrel{\raisebox{-
.6ex}{$\stackrel{\textstyle>}{\sim}$}}}

\newcommand{\drawsquare}[2]{\hbox{%
\rule{#2pt}{#1pt}\hskip-#2pt%  left vertical
\rule{#1pt}{#2pt}\hskip-#1pt%  lower horizontal
\rule[#1pt]{#1pt}{#2pt}}\rule[#1pt]{#2pt}{#2pt}\hskip-#2pt%  upper horizontal
\rule{#2pt}{#1pt}}% right vertical
\newcommand{\fund}{\raisebox{-.5pt}{\drawsquare{6.5}{0.4}}}%  fund
\newcommand{\asym}{\raisebox{-3.5pt}{\drawsquare{6.5}{0.4}}\hskip-6.9pt%
        \raisebox{3pt}{\drawsquare{6.5}{0.4}}}%  antisymmetric second rank

\begin{document}

\title{On the Unification of Gauge Symmetries in Theories with Dynamical
Symmetry Breaking}

\author{Neil D. Christensen}
%\thanks{neil.christensen@sunysb.edu}

\author{Robert Shrock}
%\thanks{robert.shrock@sunysb.edu}

\affiliation{
C.N. Yang Institute for Theoretical Physics \\
State University of New York \\
Stony Brook, NY 11794}

\begin{abstract}

We analyze approaches to the partial or complete unification of gauge
symmetries in theories with dynamical symmetry breaking. Several types of
models are considered, including those that (i) involve sufficient unification
to quantize electric charge, (ii) attempt to unify the three standard-model
gauge interactions in a simple Lie group that forms a direct product with an
extended technicolor group, and, most ambitiously, (iii) attempt to unify the
standard-model gauge interactions with (extended) technicolor in a simple
Lie group.

\end{abstract}


%\keywords{technicolor}
\maketitle

\section{Introduction}
\label{section1}

The standard model (SM) gauge group $G_{SM} = {\rm SU}(3)_c \times {\rm
SU}(2)_w \times {\rm U}(1)_Y$ has provided a successful description of both
strong and electroweak interactions.  Although the standard model itself
predicts zero neutrino masses, its fermion content can be augmented to
incorporate neutrino masses and lepton mixing.  One of the great triumphs of
this model was its unification of weak and electromagnetic interactions.
However, as has long been recognized, there are a number of properties that
this model does not explain, including the quantization of the electric charges
of elementary particles, the ratios of the values of the respective
standard-model gauge couplings $g_3$, $g_{2L}$, and $g_Y$, and the
interconnected manner in which quark and lepton contributions to gauge
anomalies cancel each other (separately for each generation).  These
deficiencies motivated the effort to construct theories with higher
unification of gauge symmetries.  Almost all of the work toward this goal 
started with the standard model, including its Higgs mechanism, and
subsequently, supersymmetric extensions of this model.

In this paper we shall investigate various approaches to the partial or
complete unification of gauge symmetries from a different viewpoint,
incorporating the standard-model gauge group but removing the Higgs mechanism
of this model and replacing it with ingredients that can produce dynamical
electroweak symmetry breaking (EWSB) and also dynamical breaking of higher
gauge symmetries.  The origin of electroweak symmetry breaking is one of the
most important outstanding questions in particle physics, and dynamical EWSB
remains an interesting alternate to the Higgs approach.  We shall focus, in
particular, on models in which electroweak symmetry breaking is due to the
formation of a bilinear condensate of new fermions interacting via an
asymptotically free, vectorial gauge interaction, generically denoted
technicolor (TC) \cite{tc,dsb}, that becomes strong at a scale $\Lambda_{TC}
\sim 300$ GeV. To communicate the electroweak symmetry breaking to the quarks
and leptons (which are technisinglets) and generate masses for these fermions,
one adds to the technicolor theory additional gauge degrees of freedom that
transform technifermions into standard-model fermions and vice versa
\cite{etc,etcrev}.  These are denoted extended technicolor (ETC) gauge bosons.

Here we shall consider several types of unification of gauge symmetries with
dynamical symmetry breaking: 
\begin{enumerate}

\item 

Models that involve sufficient unification to quantize electric charge without
embedding all of the three factor groups of the standard model in a
(semi)simple Lie group \cite{simple}.  Since $Q = T_{3L} + Y/2$ in the standard
model, a sufficient condition for this quantization is that the weak
hypercharge $Y$ be expressed as a linear combination of generators of
nonabelian gauge groups.

\item 

Models that attempt to unify the three standard-model gauge interactions in a
simple grand unified (GU) group $G_{GU}$, 
\beq
G_{GU} \supset G_{SM} 
\label{gugsm}
\eeq
and then combine this in a direct product, $G_{ETC} \times G_{GU}$ with the ETC
gauge group, $G_{ETC}$. A successful model of this type would explain charge
quantization and the relative sizes of SM gauge couplings (but not the relative
size of the SM and ETC gauge couplings).

\item 

Most ambitiously, models that attempt to unify the three SM gauge interactions
together with technicolor, or a larger gauge symmetry described by a group
$G_{SC} \supseteq G_{TC}$, in a simple Lie group $G$, 
\beq
G \supset G_{SC} \times G_{GU} \ . 
\label{scxgut}
\eeq

\end{enumerate}

In the models that we consider of types (1) and (2) above, the technicolor
group $G_{TC}$ is embedded in a larger, extended technicolor group, $G_{ETC}$:
$G_{TC} \subset G_{ETC}$.  As indicated by this subgroup relation, in these
types of models, the infinitesimal generators of the Lie algebra of $G_{TC}$
close upon themselves, as do the generators of the Lie algebra of $G_{ETC}$.
Furthermore, in these models $[G_{ETC},G_{SM}]=0$, so that the ETC gauge bosons
do not carry any SM quantum numbers.  In contrast, in models of type (3),
although $[G_{TC},G_{SM}]=0$, the commutators of the (linear combinations of)
generators of the Lie algebra of $G$ that transform technicolor indices to
SU(3)$_c$ and SU(2)$_L$ indices and vice versa (corresponding to the ETC gauge
bosons) generate the full Lie algebra of $G$, so that there is no ETC group,
{\it per se}, that forms a subgroup of $G$ that is smaller than $G$ itself.  In
these models these ETC gauge bosons generically do carry SM quantum numbers and
the corresponding generators of $G$ do not commute with the generators of
$G_{SM}$.  As will be seen below, the origin of standard-model fermion
generations is different in models of types (1) and (2), on the one hand, and
models of type (3) on the other.

The present paper is organized as follows.  In section II we review some
relevant properties of technicolor and extended technicolor.  In Section III we
consider a partial unification model of type (1) in which charge is quantized
and all symmetry breaking is dynamical, and we address the question of how one
might try to unify this model further. Sections IV and V contain analyses and
critical assessments of models of type (2) and (3).  In Section VI we discuss
the issue of how to break a unified gauge symmetry dynamically.  Section V
contains some concluding remarks and ideas for further work.  Certain general
formulas that are used throughout the paper are contained in an appendix.

\section{Technicolor and Extended Technicolor Models}
\label{section2}

In this section we discuss some relevant properties of technicolor and extended
technicolor theories.  We take the technicolor group to be $G_{TC} = {\rm
SU}(N_{TC})$.  For models of types (1) and (2), the technifermions will
comprise a standard-model family, i.e., they transform according to the
following representations of $G_{SM}$:
\beqs
& & Q_L^{a i} = {U^{a i} \choose D^{a i}}_L: \ (N_{TC},3,2)_{1/3,L}; 
\cr\cr
& & U^{a i}_R: (N_{TC},3,1)_{4/3,R}, 
\quad D^{a i}_R: (N_{TC},3,1)_{-2/3,R} 
\cr\cr
& & L^i_L = {N^i \choose E^i}_L: \ (N_{TC},1,2)_{-1,L} \cr\cr
& & \{ N^i_R \}: \ (N_{TC},1,1)_{0,R}, 
\quad E^i_R: \ (N_{TC},1,1)_{-2,R}
\cr\cr
& & 
\label{tcfamily}
\eeqs
where here the indices $a$ and $i$ are color and technicolor indices,
respectively, and the numerical subscripts refer to weak hypercharge.  Since
the $N^i_R$ are SM-singlets and the group SU(2) is free of anomalies in gauged
currents, it follows that if $N_{TC}=2$, then there can be more than just one
$N^i_R$; we indicate this by the brackets and denote this number $N_{N_R}$.  In
models of type (3) we will encounter different types of technicolor fermion
sectors.

For models of types (1) and (2), which have a well-defined ETC gauge group, we
take this group to be
\beq
G_{ETC} = {\rm SU}(N_{ETC}) \ . 
\label{getc}
\eeq
For these models, a natural procedure in constructing the ETC theory is to
gauge the generation index, assigning the first $N_{gen.}$ components of a
fundamental representation of SU($N_{ETC}$) to be the standard-model fermions
of these three generations, followed by $N_{TC}$ components which are the
technifermions with the same standard-model quantum numbers.  The fact that,
{\it a priori}, the value of $N_{gen.}$ is arbitrary except for the requirement
that the ETC theory be asymptotically free, distinguishes these types of models
 from models of type (3), where the origin and number of SM fermion generations
are more highly constrained.  Thus, models of types (1) and (2) can
automatically accomodate the observed value of SM fermion generations,
$N_{gen.}=3$, whereas, in contrast, this success is not guaranteed for a
particular model of type (3).  Given the way in which models of types (1) and
(2) gauge the generational index, it follows that for these models 
\beqs
N_{ETC} & = &N_{gen.}+N_{TC} = 3 + N_{TC} \ . 
\label{nrel}
\eeqs

The relation (\ref{nrel}) and the requirement that $N_{TC} \ge 2$ for a
nontrivial nonabelian SU($N_{TC}$) group (as required for asymptotic freedom)
together imply that $N_{ETC} \ge 5$ for models of types (1) and (2).  The
minimal choice, $N_{TC}=2$ and hence $N_{ETC}=5$, has been used for a number of
recent studies of ETC models \cite{at94}-\cite{kt}. The choice $N_{TC}=2$ is
motivated for a number of reasons; (a) with the one-family structure of
eq. (\ref{tcfamily}), amounting to $N_{TF} = 2(N_c+1)=8$ vectorially coupled
technifermions in the fundamental representation of SU(2)$_{TC}$, it can yield
an approximate infrared fixed point and associated slow running (``walking'')
of the TC gauge coupling \cite{wtc} from $\Lambda_{TC}$ up to an ETC scale
\cite{wtc}, (b) it minimizes the technicolor contributions to the electroweak
$S$ parameter \cite{scalc}, and (c) it makes possible a mechanism to account
for light neutrinos in an extended technicolor context \cite{nt,lrs}.  Although
the value $N_{TC}=2$ is thus favored, we often shall let $N_{TC}$ be arbitrary
in the present paper (subject to the requirement of asymptotic freedom of the
TC and ETC theories) in order to show the generality of certain results.

The condition $[G_{ETC},G_{SM}]=0$ in models of type (2) means that
all components of a given representation of $G_{SM}$ transform according to the
same representation of $G_{ETC}$ and all components of a given representation
of $G_{ETC}$ transform according to the same representation of $G_{SM}$.
However, this does not imply that all of the representations of $G_{SM}$
transform according to the same representation of $G_{ETC}$.  For example, in
Ref. \cite{kt} we studied a class of ETC models in which $[G_{ETC},G_{SM}]=0$
and the left-handed and right-handed representations of the charge $Q=-1/3$
quarks and techniquarks transform according to relatively conjugate
representations of $G_{ETC}$ (which were the fundamental and conjugate
fundamental representations), and similarly with the charged leptons and
technileptons, while the charge $Q=2/3$ quarks and techniquarks of both
chiralities transformed according to the same representations of $G_{ETC}$. 

ETC models of this type (2) can be classified further according to whether (i)
the ETC gauge interactions are vectorial on the SM quarks and charged leptons,
or (ii) some ETC gauge interactions are chiral on these SM fermions.  In both
cases, the TC interaction must be vectorial; this is automatically satisfied by
the SU(2)$_{TC}$ group since it has only (pseudo)real representations. (In this
SU(2)$_{TC}$ case, the number of chiral doublets is $15+N_{N_R}$, and this must
be even to avoid a global SU(2) anomaly, so $N_{N_R}$ must be odd.)  These two
options (i) and (ii) were labelled VSM and CSM in Ref. \cite{ckm}, where the V
and C referred to the corresponding vectorial and (relatively) conjugate ETC
representations of the SM quarks and charged leptons, respectively.  While both
of these classes of models have promising features, neither is fully realistic.
In Ref. \cite{ckm} it was shown that constraints from neutral flavor-changing
current processes were not as severe for VSM-type models as had been previously
thought.  However, these models require additional ingredients to produce mass
splittings within each generation, such as $m_t >> m_b, m_\tau$ (without
excessive contributions to the parameter $\rho = m_W^2/(m_Z^2 \sin^2
\theta_W)$).  The CSM-type models in which charge $-1/3$ quarks and leptons of
opposite chiralities transform according to relatively conjugate
representations of SU(5)$_{ETC}$ while the charge $2/3$ quarks have vectorial
ETC couplings can produce these requisite intragenerational mass splittings and
also some CKM mixing; however, these models do have problems with
flavor-changing neutral current processes.  Hence, we shall concentrate on
VSM-type ETC models here.  Moreover, additional ingredients are necessary to
avoid overly light Nambu-Goldstone bosons.

As an illustration, for VSM models of type (2) the fermions with SM quantum
numbers are assigned to representations of the group ${\rm SU}(5)_{ETC} \times
G_{SM}$ which are obtained from those listed in eq. (\ref{tcfamily}) by letting
the index $i$ range over the full set of ETC indices, $i=1,...,N_{ETC}$.  Here
and below, it will often be convenient to use the compact notation $Q_L$,
$L_L$, $u_R$, $d_R$, and $e_R$ to denote these ETC multiplets, so that, for
example, $e_R$, written out explicitly, is
\beq
e_R \equiv (e^1,e^2,e^3,e^4,e^5)_R \equiv (e,\mu,\tau,E^4,E^5)_R \ .
\label{er}
\eeq

Since the fermion content of the SU(5)$_{ETC}$ theory is chosen so that it is
asymptotically free \cite{b0cases}, as the energy scale decreases from large
values, the ETC coupling increases in strength.  Eventually, this coupling
becomes large enough to produce fermion condensates, and the ETC sector is
constructed to be a chiral gauge theory, so that a bilinear fermion condensate
generically self-breaks the ETC gauge symmetry. In order to obtain the desired
sequential breaking of SU(5)$_{ETC}$ to SU(2)$_{TC}$,
Refs. \cite{at94}-\cite{kt} incorporated an additional asymptotically free
gauge interaction which becomes strongly coupled at roughly the same energy
scale as the ETC interaction.  This additional interaction was called
``hypercolor'' (HC) and the corresponding gauge group was chosen to be
SU(2)$_{HC}$. The symmetry breaking of SU(5)$_{ETC}$ occurs as a combination of
self-breaking and couplings to the auxiliary strongly interacting group,
SU(2)$_{HC}$.  The fermions involved in producing the ETC symmetry-breaking
condensates are nonsinglets under the ETC group and are singlets under the
standard-model group; they include both singlets and nonsinglets under the HC
group.  The determination of which condensation channels are dynamically
favored is a difficult, nonperturbative problem involving strong coupling.  The
procedure makes use of the ``most attractive channel'' (MAC) criterion
\cite{rds} and tools such as approximate solutions of the Schwinger-Dyson
equation for the relevant fermion propagator \cite{gap}-\cite{tcvac} (see
appendix). As the energy scale decreases, the first breaking occurs at a scale
denoted $\Lambda_1$, where ${\rm SU}(5)_{ETC} \to {\rm SU}(4)_{ETC}$; here the
first generation fermions split off from the rest in each ETC multiplet.  Since
this is the highest ETC symmetry-breaking scale, we shall label it more
generally as $\Lambda_{ETC,max}$.  Similarly, one has the successive breakings
${\rm SU}(4)_{ETC} \to {\rm SU}(3)_{ETC}$ at $\Lambda_2$, where the
second-generation fermions split off, and ${\rm SU}(3) \to {\rm SU}(2)_{TC}$ at
$\Lambda_3$, where the third-generation fermions split off, leaving the exact
residual technicolor gauge symmetry.  This can account, at least approximately,
for the observed fermion masses while satisfying other constraints such as
those from flavor-changing neutral current processes, if one takes $\Lambda_1
\simeq 10^3$ TeV, $\Lambda_2 \simeq 10^2$ TeV, and $\Lambda_3 \simeq 4$ TeV, as
in Refs. \cite{at94}-\cite{kt}.  The generational ETC scales are bounded above
by the requirement that the resultant SM quark and lepton masses of the $j$'th
generation, $m_{f_j} \sim \eta_j \Lambda_{TC}^3/\Lambda_j^2$ be sufficiently
large (where $\eta_j$ is an enhancement factor present in theories with walking
technicolor \cite{wtc} and can be of order $\Lambda_3/\Lambda_{TC}$.)

An important general feature of ETC theories, illustrated in the specific
models mentioned above, is that the ETC symmetry-breaking scales are far below
the conventional grand unification scale $M_{GU} \sim 10^{16}$ GeV.  Recall
that, before engaging in detailed calculations of gauge coupling evolution, one
knows that it would be difficult to have a generic grand unification scale
lower than $10^{15} - 10^{16}$ GeV without producing excessively rapid nucleon
decay.  Therefore, insofar as one studies the possibility of unifying
standard-model and technicolor gauge symmetries, it is not technicolor itself,
but rather the higher symmetry associated with extended technicolor, that
enters into this unification.  That is, one must take account of the fact that
the effective theory at energy scales far below the grand unification scale is
already invariant under a larger symmetry involving gauge degrees of freedom
transforming technicolor indices to standard-model (color and electroweak)
indices.

\section{Example of Partial Unification of SM Gauge Symmetries and Attempt at
Higher Unification}
\label{section3}

In this section we consider the model of type (1) from Ref. \cite{lrs,nag},
which successfully achieves the important goals of electric charge quantization
via partial unification of standard-model gauge interactions with all symmetry
breaking dynamical, and we investigate how one might try to unify it further to
be a model of type (2).  A sufficient condition for the quantization of
electric charge $Q$ (or equivalently, the quantization of weak hypercharge $Y$,
given the $Q=T_{3L}+Y/2$ relation) is that $Q$ be expressed as a linear
combination of generators of nonabelian gauge groups.  An early realization of
this condition was provided by the Pati-Salam unification of color SU(3)$_c$
with U(1)$_{B-L}$ in SU(4)$_{PS}$, where $B$ and $L$ denote baryon and lepton
number \cite{ps}.  In this type of theory, the strong and electroweak gauge
groups are enlarged to the group
\beq
G_{422} = {\rm SU}(4)_{PS} \times {\rm SU}(2)_L \times {\rm SU}(2)_R \ . 
\label{g422}
\eeq
where ${\rm SU(2)}_L \equiv {\rm SU}(2)_w$ of the SM.
The left- and right-handed SM fermions of each generation are assigned to the
representations
\beq
 \left( \begin{array}{cc}
    u^{ia} & \nu^i  \\
    d^{ia} & e^i  \end{array} \right )_\chi \ , \quad \chi=L,R \ , 
\label{psrep}
\eeq
transforming as (4,2,1) and (4,1,2), respectively, under the group $G_{422}$.
The superscripts $a$ and $i$ in eq. (\ref{psrep}) refer, as before, to color
and generation.  This fermion content thus requires the addition of three
right-handed neutrinos to the standard model.  The SU(4)$_{PS}$ gauge symmetry
is vectorial, while the SU(2)$_L$ and SU(2)$_R$ symmetries are chiral.  The
electric charge operator is given by
\beqs
Q & = & T_{3L}+T_{3R}+(1/2)(B-L) \cr\cr
  & = & T_{3L}+T_{3R}+(2/3)^{1/2} \ T_{PS,15} \cr\cr
  & = & T_{3L}+T_{3R}+ (1/6) \ {\rm diag}(1,1,1,-3) \ , 
\label{qps}
\eeqs
where $T_{PS,15}=(2\sqrt{6})^{-1}{\rm diag}(1,1,1,-3)$ is the third diagonal
generator in the SU(4)$_{PS}$ Lie algebra.  The (quantized) hypercharge
generator is $Y=T_{3R}+(2/3)^{1/2}T_{PS,15}$.  Since we will analyze the
question of unification for the gauge couplings of the group $G_{422}$, we show
their explicit normalization via the covariant derivative ,
\beqs
D_\mu & = &\partial_\mu - ig_{PS} {\bf T}_{PS} \cdot {\bf A}_{PS,\mu} \cr\cr
      & &  -ig_{2L} {\bf T}_L \cdot {\bf A}_{L,\mu} - 
ig_{2R} {\bf T}_R \cdot {\bf A}_{R,\mu} \ . 
\label{dg422}
\eeqs

The model of Refs. \cite{lrs,nag} uses the gauge group
\beq
{\rm SU}(5)_{ETC} \times {\rm SU}(2)_{HC} \times G_{422}
\label{g422etc}
\eeq
with the fermion representations 
\beq
(5,1,4,2,1)_L \ , \quad (5,1,4,1,2)_R \ . 
\label{qlps}
\eeq
(This model also contains fermions that are singlets under $G_{422}$.)  In
addition to the successful quantization of electric charge and partial
unification of quarks with leptons (and techniquarks with technileptons), the
SU(4)$_{PS}$ gauge interactions connecting quarks and leptons gives mass to the
$P^0$ and $P^3$ Nambu-Goldstone bosons corresponding to the generators $I_{2V}
\times T_{PS,15}$ and $(T_3)_{2V} \times T_{PS,15}$ \cite{binsik}, where $I$
denotes the identity and the subscript $2V$ refers to vectorial isospin,
SU(2)$_V$.

In the model of Ref. \cite{lrs}, as the energy decreases below a scale
$\Lambda_{PS} \gsim \Lambda_1 \simeq 10^6$ GeV, the $G_{422}$ gauge symmetry is
broken to $G_{SM}$ by the formation of a bilinear fermion condensate.  This 
value for $\Lambda_{PS}$ satisfies experimental constraints such as those from
upper limits on right-handed charged weak currents and on the branching ratio
for the decays $K_L \to \mu^\pm e^\mp$. The matching
relations for the (running) coupling constants at this scale $\Lambda_{PS}$ are
$g_3 = g_{PS}$ and \cite{lrs} 
\beq
\frac{1}{g_Y^2} = \frac{1}{g_{2R}^2} + \frac{2}{3g_{PS}^2} \ . 
\label{gyrelps}
\eeq
Further symmetry breaking at lower scales is the same as in the
model with gauge group (\ref{getc}).  The relation between the 
electromagnetic coupling $e$ and the above gauge couplings is 
\beq
\frac{1}{e^2} = \frac{1}{g_{2L}^2} + \frac{1}{g_Y^2} = 
 \frac{1}{g_{2L}^2} + \frac{1}{g_{2R}^2} + \frac{2}{3g_{PS}^2}
\label{erel}
\eeq
 from which one can calculate the weak mixing angle 
$\sin^2 \theta_W = e^2/g_{2L}^2$ as 
\beq
\sin^2 \theta_W = \biggl [ 1 + \frac{g_{2L}^2}{g_{2R}^2}
+ \frac{2g_{2L}^2}{3g_{PS}^2} \biggr ]^{-1} \ . 
\label{swsq}
\eeq
This partial gauge coupling unification is consistent with the precision
determination of the three SM gauge couplings \cite{lrs}. Evolving the SM
couplings from $m_Z$ to $\Lambda_{PS}$, one finds, at the latter scale, the
values $\alpha_3=0.064$, $\alpha_{2L}=0.032$, and $\alpha_{PS}=0.008$ (where
$\alpha_j = g_j^2/(4\pi)$), so that the matching equations can be satisfied
with $\alpha_{2R}(\Lambda_{PS}) \simeq 0.013$, i.e., $g_{2R}/g_{2L} \simeq
0.64$ at this scale.

The results of Refs. \cite{lrs,nag} motivate one to investigate the possibility
of embedding the three factor groups of $G_{422}$ in a simple group $G_{GU}$,
which could form a direct product with $G_{ETC}$ (and possible other groups
such as an auxiliary hypercolor group).  This would promote this model of type
(1) to a model of type (2).  We observe that ${\rm SU}(4) \approx {\rm SO}(6)$
and ${\rm SU}(2) \times {\rm SU}(2) \approx {\rm SO}(4)$, so SO(10) contains,
as a maximal subgroup, the direct product ${\rm SO}(6) \times {\rm SO}(4)$.
Similarly, there is a natural embedding of the three-fold direct product
$G_{422}$ as a maximal subgroup in SO(10).  A necessary condition for this
unification is that the three gauge couplings, $g_{PS}$, $g_{2L}$, and $g_{2R}$
could plausibly evolve, as the energy scale increases, to the single SO(10)
coupling $g$.  For an exploratory study of the feasibility of this, it will be
sufficient to use one-loop renormalization group evolution equations.  Some
relevant general formulas are listed in the appendix.  We need the leading
coefficients in the beta functions for each factor group for the energy
interval above $\Lambda_{PS}$.  These are, for SU(4)$_{PS}$,
\beqs
b^{(PS)}_0 & = & \frac{1}{3}[44 - 4(N_{gen.}+N_{TC})] \cr\cr
           & = & \frac{1}{3}(32 - 4N_{TC})
\label{b4ps}
\eeqs
and, for SU(2)$_L$ and SU(2)$_R$, 
\beqs
b^{(2L)}_0 =b^{(2R)}_0 & = & \frac{1}{3}[22 -4(N_{gen.}+N_{TC})] \cr\cr
                       & = & \frac{1}{3}(10-4N_{TC}) 
\label{b2ps}
\eeqs
where the formulas are given for general $N_{TC}$ to show the fact that our
conclusions concerning unification hold for arbitrary values of this parameter.
We find that, with the fermion content as specified above, the couplings
$g_{PS}$, $g_{2L}$, and $g_{2R}$ do not unify at any higher energy scale.  In
particular, since $\alpha_{2L}$ and $\alpha_{2R}$ are unequal at $\Lambda_{PS}$
and have the same beta functions, the respective $\alpha^{-1}_{2L}$ and
$\alpha^{-1}_{2L}$ evolve as a function of $\ln \mu$ as two parallel lines,
which precludes unification.  This is still true if one augments the fermion
content of the hypothetical SO(10) theory, since the representations of SO(10)
treat the SU(2)$_L$ and SU(2)$_R$ subgroups symmetrically.  Thus, we find that
it appears to be difficult to increase the partial unification of the SM gauge
symmetries in this model to a full unification of these symmetries in a direct
product group containing ${\rm SO}(10) \times {\rm SU}(5)_{ETC}$.
Nevertheless, the model of Refs. \cite{lrs,nag} does provide an example of
partial unification of SM gauge symmetries explaining charge quantization in a
fully dynamical framework.

\section{Prospects for Models with a $G_{ETC} \times G_{GU}$ Symmetry 
Group} 
\label{section4}

\subsection{Evolution of SM Gauge Couplings in an ETC Framework} 

In this section we assess the prospects for attempting to unify the three gauge
groups of the standard model, SU(3)$_c$, SU(2)$_w$, and U(1)$_Y$, in
a simple grand unified group $G_{GU}$ which forms a direct product with the ETC
group (and possibly other groups such as hypercolor) at a high scale, $M_{GU}$.
In terms of the classification given at the beginning of the paper, these are
models of type (2).  Here and elsewhere in the paper the adjective ``grand
unified'' is used with its historical meaning, as referring to the unification
of the three SM gauge interactions only, not additional interactions such as
(extended) technicolor.  We assume that at $M_{GU}$, $G_{GU}$ breaks to the
three-fold direct product group comprising $G_{SM}$, so that in the interval of
energies extending downward from $M_{GU}$ to the highest ETC scale,
$\Lambda_{ETC,1} \simeq 10^6$ GeV, the effective field theory is invariant
under $G_{ETC} \times G_{SM}$ (times possible auxiliary groups such as
hypercolor).  As before, we take $G_{ETC} = {\rm SU}(N_{ETC})$ and, to show the
generality of our results, we keep $N_{TC}$ arbitrary (subject to the
requirement of asymptotic freedom for the ETC and TC group).

A prerequisite for this unification is the condition that the three SM gauge
couplings unify at the hypothetical scale $M_{GU}$.  The normalization of the
abelian gauge coupling is determined by the embedding of the fermions with SM
quantum numbers in the unified group, $G_{GU}$.  We shall assume that $G_{GU}$
is either of the well-known unification groups SU(5) \cite{gg} or SO(10)
\cite{guts}, with the usual assignments of SM fermions; in both cases, the U(1)
coupling that unifies with $g_3$ and $g_{2L}$ is $g_1 = \sqrt{5/3} \ g_Y$.  We
will take $\mu =m_Z$ as the starting point for the evolution of the SM
couplings to higher scales.  For our analysis, it will be adequate to use the
one-loop approximations to the respective beta functions, as given in
eqs. (\ref{beta}) and (\ref{alfsol}) of the appendix; these depend on the
leading-order coefficients $b^{(j)}_0$ for each factor group $G_j$.  It will
also be sufficient to take the top quark to be dynamical at the electroweak
scale, i.e., to include its contribution in the calculation of the beta
functions.  The values of the $b^{(j)}_0$ for the standard model with its Higgs
boson are well known: $b^{(3)}_0=(1/3)(33-2N_q)=7$,
$b^{(2)}_0=(1/3)(22-N_d-1/2)=19/6$, and $b^{(1)}_0=(3/5)b^{(Y)}_0=-41/10$,
where $N_q=2N_{gen.}=6$ denotes the number of active quarks and
$N_d=N_{gen.}(N_c+1)=12$ denotes the number of SU(2)$_w$ doublets.  It is also
well known that, if one evolves these couplings individually without further
new physics at intermediate scales, they do not unify at any one scale.

To calculate the evolution of the SM gauge couplings in the framework of an ETC
theory, we first remove the SM Higgs and, for energies above $\mu \sim
\Lambda_{TC} \sim 300$ GeV, where the technifermions are active, we add their
contributions to the $b^{(j)}_0$. The deletion of the Higgs boson from the
theory removes a term $-1/6$ from $b^{(2)}_0$, which becomes $b^{(2)}_0=10/3$,
and a term $-1/6$ from $b^{(Y)}_0$, so that $b^{(1)}_0$ becomes $b^{(1)}_0=-4$
(and leaves $b^{(3)}_0$ unchanged).  A caveat is that, even if the SM gauge
couplings are small at a given scale $\mu$, their evolution may still be
significantly affected by nonperturbative, strong couplings of the
SM-nonsinglet technifermions.  Since we start our integration of the
renormalization group equations at $\mu=m_Z$, which is comparable to, and,
indeed, slightly less than, the technicolor scale, $\Lambda_{TC} \simeq 300$
GeV, these strong technifermion interactions produce some uncertainty in the
evolution of the SM gauge couplings.  This is a consequence of the fact that in
the beta function calculations, one treats the technifermions as weakly
interacting, but this is only a good approximation for $\mu >> \Lambda_1 \simeq
10^6$ GeV.

Generalizing $N_q$ to refer to both quarks and techniquarks, and letting $N_d$
denote the total number of SU(2)$_L$ doublets, we calculate 
\beqs
b^{(3)}_0 & = & \frac{1}{3}(11N_c - 2N_q) \cr\cr
          & = & \frac{1}{3}[33 - 4(N_{gen.}+N_{TC})] \cr\cr
          & = &  7 - \frac{4}{3}N_{TC}
\label{b3_etc}
\eeqs
\beqs
b^{(2)}_0 & = & \frac{1}{3}(11N_w - N_d) \cr\cr
          & = & \frac{1}{3}[22 - (N_c+1)(N_{gen.}+N_{TC})] \cr\cr
          & = &  \frac{10}{3} - \frac{4}{3}N_{TC}
\label{b2_etc}
\eeqs
and
\beqs
b^{(1)}_0 & = & \frac{3}{5}b^{(Y)}_0 = -\frac{4}{3}(N_{gen.}+N_{TC}) \cr\cr
          & = & -4 - \frac{4}{3}N_{TC} \ . 
\label{by_etc}
\eeqs
As is evident in these results, the respective beta functions, and, in
particular, the leading coefficients $b^{(j)}_0$, depend on $N_{TC}$ only
through the combination $N_{gen.}+N_{TC}=N_{ETC}$.  Consequently, the addition
of the one family of technifermions to the fermion content of the standard
model is equivalent to the addition of $N_{TC}$ additional generations of SM
fermions.  Now we recall that the addition of one or more (complete)
generations of SM fermions leaves the differences $(\Delta b_0)_{ij} \equiv
b^{(i)}_0-b^{(j)}_0$, $ij=12,13,23$ invariant \cite{gu}. (Explicitly, $(\Delta
b_0)_{32}=11/3$, $(\Delta b_0)_{31}=11$, and $(\Delta b_0)_{21}=22/3$.)  Hence,
for a given set of values of $\alpha_j$, $j=1,2,3$ at $\mu = m_Z$, the scales
$\mu_{ij}$ where $\alpha_i=\alpha_j$ are independent of $N_{TC}$, in the same
way as these differences are independent of $N_{gen.}$.  Therefore, just as
there was no gauge coupling unification in the SM and the SM without a Higgs,
so also, this remains true for the theory with one family of technifermions
added.  For reference, we note that the scales $\mu_{ij}$ at which pairwise
equalities of couplings occur are roughly $\mu_{23} \simeq 10^{18}$ GeV,
$\mu_{13} \simeq 10^{14.5}$ GeV, and $\mu_{12} \simeq 10^{12.8}$ GeV.

\subsection{An Implication of Unification Involving $G_{ETC} \times G_{GU}$}

Even if one were able to achieve unification of the three standard-model gauge
symmetries in a simple group $G_{GU}$ which commutes with the ETC group in a
model of type (2), one would encounter another problem.  The unification groups
$G_{GU}$ that have been studied have the property that for at least one fermion
$f$, both $f_L$ and $f^c_L$ are contained in a given representation of
$G_{GU}$.  For example, in the SU(5) model of Ref. \cite{gg}, $u_L$ and
$u^c_L$ are both assigned to the rank-2 antisymmetric tensor representation,
the $10_L$.  In the SO(10) model, all of the left-handed components and
conjugates of the right-handed components of quarks and leptons are contained
in the 16-dimensional spinor representation.  The property that the full gauge
group contains the direct product $G_{ETC} \times G_{GU}$ means that
$[G_{ETC},G_{GU}]=0$.  It follows that any $f_L$ and $f^c_L$ that belong to the
same representation of $G_{GU}$ also transform according to the same
representation ${\cal R}_{ETC}$ of $G_{ETC}$, or equivalently, $f_L$ and $f_R$
transform according to relatively conjugate representations ${\cal R}_{ETC}$
and $\bar {\cal R}_{ETC}$ \cite{lrs}. Here we use $f_L$ and $f_R$ to refer to
all of the fermions of these respective chiralities with the same SM quantum
numbers, as illustrated for $e_R$ in eq. (\ref{er}). This strongly suppresses
the mass $m_{f^j}$ that is produced, for a given generation $j$.  This is true
for the $Q=2/3$ quarks ($f=u$) in the case $G_{GU} = {\rm SU}(5)_{GU}$ and for
all of the SM fermions in the case $G_{GU}= {\rm SO}(10)$.  This suppression
would render it very difficult to obtain adequate quark and lepton masses, in
particular, the top quark mass.  (Recall that the CSM models of Ref. \cite{kt}
always used vectorial ETC representations for $Q=2/3$ quarks and techniquarks.)
Moreover, there would be serious problems associated with excessive ETC
contributions to flavor-changing neutral current processes, as was shown in
Ref. \cite{kt} for the CSM case where the $Q=-1/3$ quarks and leptons of
opposite chiralities transform according to relatively conjugate
representations of $G_{ETC}$.

\section{On the Unification of SM and TC Gauge Symmetries in a Simple Group} 
\label{section5}

\subsection{General}

Here we consider models of type (3), which attempt to achieve the unification
of the three standard-model gauge symmetries with a group $G_{SC}$ which
contains technicolor,
\beq
G_{SC} \supseteq G_{TC} \ , 
\label{gsctc}
\eeq
and possibly also some generational symmetries, in a unified gauge symmetry
described by the group, $G$, as specified in eq. (\ref{scxgut}). The physics is
invariant under the symmetry group $G$ at energies above the unification scale,
$M_{GU}$, and this symmetry breaks at $M_{GU}$.  The subscript $SC$ indicates
that the $G_{SC}$ gauge interaction becomes strongly coupled at a scale which
is roughly comparable to conventional ETC scales, although it is small and
perturbative at the high scale $M_{GU}$. These are the most ambitious of all of
the three types of models considered here.  They entail the unification of the
three SM gauge couplings and the SC gauge coupling at $M_{GU}$.  Thus, the
absence of this gauge coupling unification would, by itself, be enough to
exclude such models.  However, the analysis of the evolution of the relevant
gauge couplings is more complicated for these models because of the
nonperturbative behavior and associated dynamical symmetry breaking that occurs
at intermediate energy scales between $m_Z$ and the hypothesized $M_{GU}$; as a
consequence of this, one cannot use the perturbative evolution equation
(\ref{alfsol}) for all of the relevant couplings.  We will discuss this further
below.

   As soon as one hypothesizes a technicolor gauge symmetry as a dynamical
mechanism for electroweak symmetry breaking, it is natural to explore the idea
of trying to unify technicolor with the three SM factor groups - color, weak
isospin, and weak hypercharge - in a simple group.  The motivations for this
are similar to the motivations for the original grand unification program,
including a unified description of the fermion representations and an
explanation of the relative coupling strengths at lower energies, including, in
particular, the property that the TC interaction becomes strongly coupled at a
scale $\Lambda_{TC} \simeq 300$ GeV, which is essentially the electroweak
scale, and which is about $10^3$ times larger than the scale $\Lambda_{QCD}
\simeq 0.2$ GeV at which the color coupling becomes large.  In this approach,
one would, {\it a priori}, plausibly hope to explain the large ratio
$\Lambda_{TC}/\Lambda_{QCD}$ and hence also $m_Z/\Lambda_{QCD}$, as a
consequence of moderate differences in the relevant beta functions, together
with the property of slow, logarithmic running over the energy interval between
$M_{GU}$ and the highest scale at which some couplings grow to be of order
unity.  Early studies on the possibility of grand unification of technicolor
and standard-model symmetries were \cite{fs,fr}.

Perhaps the simplest notion of unification of technicolor with the three
standard-model gauge interactions would be to have a simple gauge group $G$
that contains all four of these interactions as subgroups, $G \supset
G_{TCSM}$, where $_{TCSM} = G_{TC} \times G_{SM}$, such that $G$ breaks to
$G_{TCSM}$ at a high scale $M_{GU}$.  This hypothetical theory would be
constructed so that the technicolor beta function would be more negative than
the SU(3)$_c$ beta function, $\beta_{TC} < \beta_{SU(3)_c} < 0$, and hence, as
the energy scale decreases, the technicolor gauge coupling would become
sufficiently large to cause a technifermion condensate at a scale
$\Lambda_{TC}$ well above the scale $\Lambda_{QCD}$ at which the SU(3)$_c$
coupling gets large and produces the $\langle \bar q q \rangle$ condensate.
However, this approach is excluded immediately by the fact that the gauge
bosons in $G$ that transform technifermions into the technisinglet
standard-model fermions and hence communicate the electroweak symmetry breaking
to the latter and give them masses lie in the coset $G/G_{TCSM}$ and hence pick
up masses of order $M_{GU}$.  The effective ETC scale would thus be the grand
unification scale, $M_{GU}$, resulting in standard-model fermion masses that
are much too small; for example, for the illustrative value $M_{GU} = 10^{16}$
GeV, these fermion masses would be of order $\Lambda_{TC}^3/M_{GU}^2 \simeq
10^{-25}$ GeV.  It should be noted that this early approach to the unification
of technicolor and SM gauge symmetries led to the inference that $N_{TC}$ had
to be greater than $N_c=3$.  But since this attempt at unification is
immediately ruled out by its failure to obtain fermion masses of adequate size,
its requirement concerning $N_{TC}$ is only of historical interest, and,
indeed, many recent TC models \cite{at94}-\cite{kt} use $N_{TC}=2$ for the
reasons that we have discussed in Section II \cite{bcond}.

Here we consider a different approach to this goal of unification, in which the
ETC gauge bosons have masses in the usual ETC range, and not all of the fermion
generations arise from the representations of the unified group $G$ but
instead, some arise from sequential symmetry breaking of a smaller subgroup of
$G$ at ETC-type scales.  Let us denote $N_{gh}$ and $N_{g \ell}$ as the numbers
of standard-model fermion generations arising from these two sources,
respectively, where the subscripts $gh$ and $g \ell$ refer to
\underline{g}enerations from the representation content of the
\underline{h}igh-scale symmetry group and from the \underline{l}ower-scale
breaking.  Together, these equal the observed number of SM fermion generations:
\beq
N_{gen.} = 3 = N_{gh} + N_{g \ell} \ . 
\label{ngen} 
\eeq
Note that in this approach involving sequential symmetry breaking, one does not
calculate the beta functions of the low-energy SC or TC sectors by enumerating
the fermion content at the unification scale since some subset of these
fermions would be involved in condensates formed at intermediate energy scales,
hence would gain dynamical masses of order these scales, and would be
integrated out before the energy decreases to the scale relevant for the
evolution of SC or TC gauge couplings. It should also be remarked that at this
stage the number $N_{g \ell}$ is only formal; that is, we set up a given model
so that, {\it a priori}, it can have the possibility that a subgroup of $G$
such as $G_{SC}$ might break in such a manner as to peel off $N_{g \ell}$ SM
fermion generations.  In fact, we will show that, at least in the models
that we study, it is very difficult to arrange that this desired symmetry
breaking actually takes place.

The requirement that the ETC gauge bosons have masses of the necessary scales
means that $G$ cannot break to the direct product group $G_{TCSM}$ at the
unification scale $M_{GU}$, and also cannot break at this scale to the larger
subgroup $_{SCSM}=G_{SC} \times G_{SM}$.  Instead, $G$ must break to a direct
product group such that one or more of the factor groups that are residual
symmetries between $M_{GU}$ and $\Lambda_{ETC,max}$ contain gauge bosons that
transform technifermions into technisinglet standard-model fermions, i.e. are
ETC gauge bosons.  As the energy scale decreases, this intermediate symmetry
should break at $\Lambda_{ETC,max}$ so that some of the ETC gauge bosons get
masses of this order, and so forth for other lower sequential ETC scales.

To provide an explicit context for our analysis, let us consider unifying the
SU($N_{SC})$ symmetry containing technicolor with the SM gauge symmetries by
using a group $G={\rm SU}(N)$ as in eq. (\ref{scxgut}), with
\beq
N = N_{SC} + N_c + N_w = N_{SC} + 5 \ . 
\label{n}
\eeq
Thus, $G \supset G_{SCSM}$.  Here we shall take $G_{TC}={\rm SU}(N_{TC})$,
$G_{SC}={\rm SU}(N_{SC})$, and $G_{GU}$ to be the group SU(5)$_{GU}$ of
Ref. \cite{gg}.  The fermion representations are determined by the structure of
the fundamental representation, which we take to be
\beq
\psi_R = \left( \begin{array}{c}
        (N^c)^\tau \\
         d^a \\
        -e^c \\
        \nu^c \end{array} \right )_R
\label{5genR}
\eeq
where $d$, $e$, and $\nu$ are generic symbols for the fermions with these
quantum numbers.  Thus, the indices on $\psi_R$ are ordered so that the indices
in the SC set, which we shall denote $\tau$, take on the values
$\tau=1,...N_{SC}$ and then the remaining five indices are those of the $5_R$
of SU(5)$_{GU}$ \cite{gg}. The components of $N^c_R$ transform according to the
fundamental representation of SU($N_{SC}$), are singlets under SU(3)$_c$ and
SU(2)$_w$, and have zero weak hypercharge (hence also zero electric charge).
Our choice to write these components as $(N^c)^\tau_R$ instead of $N^\tau_R$ is
a convention.  The quantum numbers of components of any representation of $G$
are determined by the structure of the fundamental representation
(\ref{5genR}).  This structure is concordant with the direct product in 
eq. (\ref{scxgut}) and the corresponding commutativity property
\beq
[G_{SC}, G_{GU}]= 0
\label{gsccomm}
\eeq
which, since $G_{SC} \supseteq G_{TC}$, implies
\beq
[G_{TC}, G_{GU}]= 0 \ . 
\label{gtccomm}
\eeq

These properties have important consequences for fermion masses.  We recall the
theorem from Ref.  \cite{lrs} discussed in Section IVB, that
$[G_{ETC},G_{GU}]=0$ implies that for one or more fermions $f$, since $f_L$ and
$f^c_L$ are both contained in the same representation of $G_{GU}$, $f_L$ and
$f_R$ transform according to relatively conjugate representations of $G_{ETC}$.
By the same argument, the commutativity property (\ref{gsccomm}) implies that
for one or more fermions $f$, since $f_L$ and $f^c_L$ are both contained in the
same representation of $G_{GU}$, $f_L$ and $f_R$ transform according to
relatively conjugate representations of $G_{SC}$ and hence also of $G_{TC}$.
For the case $G_{GU}={\rm SU}(5)_{GU}$ on which we focus here, this includes
the charge 2/3 techniquarks.  In the models of types (1) and (2) discussed in
Section IVB, this would lead to the strong suppression of the masses of the
TC-singlet SM fermions with these quantum numbers, i.e., the charge 2/3 quarks.
Here, it will also lead to the strong suppression of certain SM fermion masses,
but because the ETC vector bosons carry color and charge in the present
type-(3) models, the fermions with suppressed masses will be leptons.

Corresponding to the subgroup decomposition $G \supset G_{SC} \times G_{GU}$,
the Lie algebra of $G$ contains subalgebras for $G_{SC}$ and $G_{GU}$. There
are also (linear combinations of) generators of $G$ that transform the $N_{SC}$
indices $\tau=1,...N_{SC}$ to the last five indices in the fundamental
representation, and vice versa.  The gauge bosons corresponding to these
generators include some of the ETC gauge bosons and have nontrivial SM quantum
numbers.  We label the basic transitions as
\beq
(N^c)^\tau_R \to d^a_R + V^\tau_a
\label{via}
\eeq
\beq
(N^c)^\tau_R \to e^c_R + (U^-)^\tau
\label{uminus}
\eeq
and
\beq
(N^c)^\tau_R \to \nu^c_R + (U^0)^\tau
\label{uzero}
\eeq
where $\tau=1,...,N_{SC}$, and the $V^\tau_a$ and ${U^0 \choose U^-}^\tau$ are
ETC gauge bosons.  Under the group $G_{SCSM}$ these transform according to
\beq
V^\tau_a \ : \quad (N_{SC},\bar 3, 1)_{2/3} \ , \quad Q=1/3
\label{viarep}
\eeq
and
\beq
{U^0 \choose U^-}^\tau \ :  \quad (N_{SC},1,2)_{-1/2} \ . 
\label{urep}
\eeq
These are thus quite different from the ETC gauge bosons of models (1) and (2),
which carry no SM quantum numbers.  It is important to note that the
(commutators of the) generators to which these ETC gauge bosons correspond do
not close to yield a subalgebra smaller than the full Lie algebra of $G$, so
that there is no ETC group $G_{ETC}$, as such.  This is analogous to the fact
that the (linear combinations of) generators that transform color indices to
electroweak indices in the conventional SU(5) grand unified theory do
not close to form an algebra smaller than the Lie algebra of SU(5) itself.
This is one of the features that distinguish these models of type (3) from the
models of types (1) and (2), which do have well-defined ETC gauge groups.

To delineate the remaining ETC gauge bosons, let us divide the SU($N_{SC}$)
indices into (a) the ones for SU($N_{TC}$), say $\tau=1,...,N_{TC}$, and (b)
the ones for the coset $G_{SC}/G_{TC}$, $\hat \tau=N_{TC}+1,...,N_{SC}$.  
Consider the process $(N^c)^\tau_R \to (N^c)^{\hat \tau}_R +
V^\tau_{\hat \tau}$; the $V^\tau_{\hat \tau}$'s constitute the remaining ETC
gauge bosons.  These carry no SM quantum numbers and are similar in this regard
to the ETC gauge bosons of models (1) and (2).

Given that $G$ cannot break completely to $G_{TCSM}$ or $G_{SCSM}$ at the
unification scale $M_{GU}$ in a viable model, we next investigate which
subgroup it could break to at this scale.  The breaking pattern must be such as
to satisfy the upper limits on the decays of protons and otherwise stable bound
neutrons.  The gauge bosons of $G_{GU}$ that contribute to these decays are the
set ${X_a \choose Y_a}$, (where $a$ is a color index) which transform as $(\bar
3,2)_{5/6}$ under ${\rm SU}(3)_c \times {\rm SU}(2)_w \times {\rm U}(1)_Y$
(whence $Q_X=4/3$, $Q_Y=1/3$), and their adjoints.  These 12 gauge bosons span
the coset ${\rm SU}(5)_{GU}/G_{SM}$ and must gain masses of order $M_{GU}$.
{\it A priori}, the breaking at $M_{GU}$ could leave an invariant subgroup
${\rm SU}(3)_c \times G_{SCW}$, where $G_{SCW}$ is a group of transformations
on the $N_{SC}$ indices of ${\rm SU}(N_{SC})$ and the two electroweak indices
of SU(2)$_w$ which naturally takes the form ${\rm SU}(N_{SC}+2)$.  However, it
appears quite difficult to construct a model of this sort because of the
conflicting requirements that the $G_{SCW}$ coupling get large, as is necessary
for self-breaking (see further below) and that the {SU}(2)$_w$ coupling, which
is supposed to have a small, perturbative value of $\alpha_2 = \alpha_{GU}$ at
the presumed unification scale, $M_{GU}$ and then evolve to its similarly
small, perturbative value of $\alpha_2(m_Z) = 0.034$ at the electroweak scale.

An alternative is that the breaking of $G$ at $M_{GU}$ would leave an invariant
subgroup ${\rm SU}(2)_w \times G_{SCC}$, where $G_{SCC}$ is a group of
transformations on the $N_{SC}$ indices of ${\rm SU}(N_{SC})$ and the three
color indices of SU(3)$_c$, which would naturally take the form ${\rm
SU}(N_{SCC})$ with 
\beq
N_{SCC} = N_{SC} + N_c = N_{SC}+3 \ . 
\label{nscc}
\eeq
Thus, 
\beq
{\rm SU}(N_{SCC}) \supset {\rm SU}(N_{SC}) \times {\rm SU}(3)_c \ . 
\label{gscc}
\eeq
All representations of SU($N_{SCC}$) are determined by the fundamental
representation, which follows directly from eq. (\ref{5genR}); again, it is
convenient to write this as a right-handed field, 
\beq
   \left( \begin{array}{c}
        (N^c)^\tau \\
         d^a \end{array} \right )_R \ . 
\label{5genRscc}
\eeq
As is evident from this, the components of a representation of ${\rm
SU}(N_{SCC})$ do not, in general, have the same weak hypercharge $Y$, so this
representation does not have a well-defined value of $Y$.  Considerations of
the relative sizes of gauge couplings would favor this option over the one
involving $G_{SCW}$ because the color SU(3)$_c$ coupling $\alpha_3$ increases
substantially from the presumed unification value at $M_{GU}$ to $\alpha_3(m_Z)
= 0.118$ at the electroweak scale.  To begin with, several possibilities could
be envisioned for the symmetry breaking of ${\rm SU}(N_{SCC})$.  One would be
that, as the energy scale decreases from $M_{GU}$, $\alpha_{SCC}$ becomes
sufficiently large at a scale $\Lambda_{ETC,max}$ for the breaking ${\rm
SU}(N_{SCC}) \to {\rm SU}(N_{SC}) \times {\rm SU}_c$ to occur, and then, if
$N_{SC} > N_{TC}$, the further sequential breakings of SU($N_{SC}$), eventually
yielding the exact symmetry group SU($N_{TC})$, peeling off the $N_{g \ell}$ SM
fermion generations.  A different scenario would be one in which, as the energy
scale decreases below $M_{GU}$, SU($N_{SCC}$) breaks into smaller simple groups
in a sequence of $N_{g \ell}$ steps, and then the residual smaller simple group
finally splits into the direct product ${\rm SU}(N_{TC}) \times {\rm SU}(3)_c$.
A third possibility would be a combination of these two types of breakings, in
which ${\rm SU}(N_{SCC})$ breaks to smaller simple groups in $k < N_{g \ell}$
stages, then splits to a two-fold direct product group one of whose factor
groups is SU(3)$_c$, and then the other factor group sequentially self-breaks
$N_{g \ell}-k$ times to the residual exact SU($N_{TC}$) group.  In all of these
scenarios, the $V^\tau_a$'s and $V^\tau_{\tau'}$ would have masses in the usual
ETC range, from $\Lambda_1$ down to $\Lambda_3$ while the $(U^0)^\tau$ and
$(U^-)^\tau$ would have masses of order $M_{GU}$.  (Hence, although
$(U^0)^\tau$ and $(U^-)^\tau$ are formally ETC gauge bosons, they would play a
negligible role in producing masses for SM fermions.)

However, this scenario with a $G_{SCC}$ gauge symmetry characterizing the
effective theory between $M_{GU}$ and $\Lambda_{ETC,max}$ also encounters a
complication. Consider, for example, the version in which SU($N_{SCC})$ would
break to ${\rm SU}(N_{SC}) \times {\rm SU}(3)_c$ at roughly $\Lambda_{ETC,max}
\simeq 10^6$ GeV.  The matching conditions for gauge couplings imply that at
this energy scale the SU(3)$_c$ and $G_{SC}$ couplings are equal, since they
inherit the value that the SU($N_{SCC})$ coupling had just above
$\Lambda_{ETC,max}$. But, assuming that the unified theory with gauge group $G$
is self-contained, i.e., there is no direct product at the high scale $M_{GU}$
with an auxiliary group like hypercolor, the only mechanism for dynamical
symmetry breaking below $M_{GU}$ is self-breaking.  This implies that at the
above energy scale of $\sim 10^6$ GeV, $\alpha_{SCC}$ should be $O(0.1)-O(1)$,
depending on the attractiveness of the relevant fermion condensation channel(s)
as measured by the respective values of $\Delta C_2$, defined in
eq. (\ref{deltac2}) in the appendix. This is difficult to reconcile with the
SU(3)$_c$ beta function, which has leading coefficient $b_0^{(3)}=7$ due to the
SM fermions, and, in the energy interval above $\Lambda_{TC}$,
$b^{(3)}_0=7-(4/3)N_{TC}$ for one family of massless technifermions.  That is,
at an energy scale of $\sim 10^6$ GeV, the value of $\alpha_3$ would already be
larger than its value at the electroweak scale, which would imply that it would
have to decrease, rather than increasing, as the scale $\mu$ decreases from
$10^6$ GeV to $\mu=m_Z$, i.e., that the SU(3)$_c$ sector would have to be
non-asymptotically free in this interval.  Another problem is that in the
models that we have constructed and studied, we find that the ${\rm
SU}(N_{SCC})$ theory is unlikely to break in the necessary manner; to explain
this, it is first necessary to describe the fermion representations, which we
do below.

Let us proceed with the construction and critical evaluation of the prospects
for this class of unified models.  The following conditions are equivalent:
\beq
G_{SC} = G_{TC} \ \Longleftrightarrow \ N_{gh}=3, \ N_{g \ell}=0 \ , 
\label{gsceqgtc}
\eeq
i.e., in this case, all of the fermion generations would arise from the fermion
representations of $G$.  The other formal possibility is that $N_{g \ell} \ge
1$ so that the coset $G_{SC}/G_{TC}$ is nontrivial, with $G_{SC}$ containing
some gauged generational structure.  For generality, it should be noted that
although we use $G_{GU}$ to classify fermion representations, the breaking of
$G$ may be such that the lower-energy effective field theory is not actually
symmetric under a direct product group in which $G_{GU}$ occurs as a factor
group.  From the property (\ref{scxgut}), it follows that the ranks satisfy 
\beq
r(G) \ge r(G_{SC})+r(G_{GU})
\label{rg}
\eeq
where, with our assumption that $G_{SC}={\rm SU}(N_{SC})$, we have
$r(G_{SC})=N_{SC}-1$.  In addition to the subgroup decomposition
(\ref{scxgut}), we will also use the subgroup decomposition 
\beq
{\rm SU}(N) \supset {\rm SU}(2)_w \times {\rm SU}(N_{SCC}) \ . 
\label{wxscc}
\eeq

We shall assume that at energies below the unification scale $M_{GU}$ all
subsequent breaking of gauge symmetries is dynamical.  Since $M_{GU}$ is not
very far below the Planck scale, which certainly constitutes an upper limit to
the possible validity of the theory, owing to the lack of inclusion of quantum
gravity, it is not clear that one needs to assume that the initial breaking of
$G$ is dynamical.  We shall comment on this further below.  The dynamical
symmetry breaking at energies below $M_{GU}$ can be classified as being of two
general types: (i) self-breaking (``tumbling''), in which an asymptotically
free chiral gauge symmetry group has an associated coupling that becomes large
enough to produce a fermion condensate that breaks the gauge symmetry, and (ii)
induced breaking in which a gauge symmetry is weakly coupled, but is broken by
the formation of condensates involving fermions that are nonsinglets under a
strongly coupled group (which is the way that electroweak symmetry is
broken by technifermion condensates); (iii) a combination of the two, as in the
sequential breaking of the SU(5)$_{ETC}$ symmetry in
Refs. \cite{at94}-\cite{kt}.  Since we only consider unification in a single,
simple group $G$ here, we are led to focus on self-breaking below $M_{GU}$.
Note that with our choice of the minimal GU group as SU(5) with rank 4, the
inequality (\ref{rg}) becomes $r(G) \ge r(G_{SC})+4$.  Our choice $G={\rm
SU}(N)$ with $N$ given by eq. (\ref{n}) satisfies this inequality as an
equality; i.e., we are choosing the minimal $G$ for a given value of $N_{SC}$.
In order to maintain the nonabelian structure of the TC group and hence the
asymptotic freedom that leads to confinement and the formation of the EWSB
bilinear fermion condensate, we require that $N_{TC} \ge 2$.  This yields the
inequality
\beq
N \ge 7
\label{nge7}
\eeq
and $r(G) \ge 6$.

For any of the possible types of sequential breakings of $G_{SCC}$ and/or
$G_{SC}$ described above that produce the $N_{g \ell}$ SM fermion generations,
one has
\beq
N_{g \ell} = N_{SCC} - (N_{TC}+N_c) \ . 
\label{ngensccrel}
\eeq
In particular, if $G_{SCC}$ first splits to $G_{SC} \times {\rm SU}(3)_c$ and 
$G_{SC}$ then sequentially breaks to produce these $N_{g \ell}$ generations,
then 
\beq
N_{g \ell} = N_{SC} - N_{TC} \ . 
\label{ngenscrel}
\eeq
The requirement that $N_{TC} \ge 2$ in order for the
technicolor interactions to be asymptotically free, 
combined with eq. (\ref{ngenscrel}), implies 
\beq
N_{g \ell} \le  N_{SC} - 2 \ . 
\label{ngenscrela}
\eeq
Thus, for $N_{SC}=2$ all of the SM fermion generations must arise via
$N_{gh}$.

We next specify the fermion representations of $G={\rm SU}(N)$.  Without loss
of generality, we shall usually deal with left-handed fermions (or
antifermions).  In order to avoid fermion representations of SU(3)$_c$ and
SU(2)$_w$ other than those experimentally observed, namely singlets and
fundamental or conjugate fundamental representations, one restricts the
fermions to lie in $k$-fold totally antisymmetrized products of the fundamental
or conjugate fundamental representation of SU($N$) \cite{g79}; we denote these
as $[k]_N$ and $[\bar k]_N = \overline{[k]}_N$.  The notational correspondence
with Young tableaux is (suppressing the dependence on $N$), $[1] \equiv \fund$,
$[2] \equiv \asym$, etc.  Some elementary properties of the representation
$[k]_N$ are listed in the appendix; these include its dimensionality and
expressions for the Casimir invariants $C_2([k]_N)$ and $T([k]_N)$. A set of
(left-handed) fermions $\{ f \}$ transforming under $G$ is thus given by
\beq
\{ f \} = \sum_{k=1}^{N-1} \ n_k \ [k]_N
\label{fermionset}
\eeq
where $n_k$ denotes the multiplicity (number of copies) of each
representation $[k]_N$.  We use a compact vector notation
\beq
{\bf n} \equiv (n_1,...,n_{N-1})_N \ .
\label{ndef}
\eeq
If $k=N-\ell$ is greater than the integral part of $N/2$, we shall work with
$[\bar \ell]_N$ rather than $[k]_N$; these are equivalent with respect to
SU($N$) (see eq. (\ref{nkn}) in the appendix).  An optional additional
constraint would be to require that the numbers in the set of $n_k$,
$k=1,...,N-1$ have no common factors greater than unity, i.e., the greatest
common divisor $GCD(\{n_k \})=1$.  This might be viewed as a kind of
irreducibility condition.  Although we will not impose this condition here, the
two candidate models that we consider that have $GCD(\{n_k \}) \ge 2$ are
excluded anyway because the SCC theory is not asymptotically free.  A fermion
field corresponding to $[k]_N$ is denoted generically by $\psi^{i_1...i_k}_L$.
It will sometimes be convenient to deal with the charge-conjugate 
right-handed field.

Before proceeding, it is appropriate to summarize the requirements on the 
choice of fermion representations:

\begin{enumerate}

\item

The theory must contain a mechanism to break the unified $G$ gauge symmetry,
eventually down to the symmetry group operative above the electroweak scale,
${\rm SU}(N_{TC}) \times G_{SM}$.  The breaking scales must be such as to obey
upper bounds on the decay rate for protons and bound neutrons.

\item 

The contributions from various fermions to the total SU($N$) gauge anomaly
must cancel each other, yielding zero gauge anomaly. 

\item 

The resultant TC-singlet, SM-nonsinglet left-handed fermions must comprise a
set of generations, i.e., must have the form $N_{gen.}[(1,\bar 5)_L+(1,10)_L]$,
where the first number in parentheses signifies that these are TC-singlets and
the second number denotes the dimension of the SU(5)$_{GU}$ representation.

\item

For a fully realistic model, one requires $N_{gen.}=3$. 

\item 

In order to account for neutrino masses, one needs to have TC-singlet,
electroweak-singlet neutrinos to produce Majorana neutrino mass terms that can
drive an appropriate seesaw \cite{nt}.  In the present context, these are also
singlets under $SU(5)_{GU}$.

\item 

The model must contain ETC gauge bosons with masses in the general range from a
few TeV to $10^3$ TeV so as to produce acceptable SM fermion masses.  As
explained above, a plausible way to satisfy this requirement is for $G$ to
break to the subgroup (\ref{wxscc}) containing the factor group $G_{SCC}$ which
contains both ${\rm SU}(N_{TC})$ and SU(3)$_c$ and is naturally SU($N_{SCC}$)
with $N_{SCC}$ given by eq. (\ref{nscc}).  Thus, the effective field theory at
energy scales between $M_{GU}$ and $\Lambda_{ETC,max} \simeq 10^6$ GeV is
invariant under this direct product (\ref{wxscc}).  The dynamics should be such
that SU($N_{SCC}$) breaks at ETC scales, in one of the ways delineated above,
eventually yielding the residal exact symmetry group ${\rm SU}(2)_{TC} \times
{\rm SU}(3)_c$, with the requisite three SM fermion generations emerging.  If
at least some of the stages of this process involve self-breaking, then the SCC
sector should be an asymptotically free chiral gauge theory. 

\item 

If $N_{SC} > N_{TC}$, then there must be a mechanism to break ${\rm
SU}(N_{SC})$ down in $N_{g \ell}$ stages to ${\rm SU}(N_{TC})$. Again, if this
is to be a self-breaking, then the SC sector should be an asymptotically free
chiral gauge theory so that the associated coupling will increase sufficiently
as the energy scale decreases to produce the requisite condensate(s).  

\item 

The color SU(3)$_c$ interaction must be asymptotically free in the
energy interval at and below the electroweak scale, where the associated
coupling has been measured.  

\item 

The technicolor interaction must asymptotically free, so that the associated
gauge coupling will increase sufficiently, as the energy scale decreases, to 
produce a technifermion condensate and break the electroweak symmetry; further,
the technicolor symmetry must be vectorial so that the technifermions are
confined and the technifermion condensate does not self-break $G_{TC}$.

\item 

When evolved down to the low energies, the respective SM gauge couplings must
agree with their measured values.  

\end{enumerate}

Let us check that the first and sixth of these constraints can be
simultaneously satisfied.  This requires that one confirm that the masses of
the (mass eigenstates corresponding to the interaction eigenstates) $V^\tau_a$
needed for their role as ETC vector bosons are consistent with the upper bounds
on the decays of protons and bound neutrons.  These decays are induced by the
$s$-channel transitions (1) $u^a + u^b \to X_c$, (2) $u^a + d^b \to Y_c$ (and,
if the theory contained another vector boson, $\Xi_c$ with $Q_\Xi=-2/3$, also
(3) $d^a + d^b \to \Xi_c$), where $a,b,c$ are (different) color indices.
Corresponding transitions in the $t$ and $u$ channel also contribute.  Among
the $G_{SC}$-nonsinglet gauge bosons, the only ones that transform in the right
way under color and electric charge to contribute to these decays are the
$V^\tau_a$, with charge 1/3.  Among these, the subset with SC indices $\tau$ in
$G_{TC}$ cannot contribute to these decays, since the $G_{TC}$ technicolor
symmetry is exact, but the quarks in a nucleon are technisinglets.  If $N_{g
\ell}=0$, then $G_{SC}=G_{TC}$, so all of the SC indices are in $G_{TC}$.  If
$N_{g \ell} \ge 1$, then there is also a subset of $V^{\hat \tau}_a$ with
indices $\hat \tau$ in the coset $G_{SC}/G_{TC}$.  The exact symmetries (color,
electric charge, and technicolor) allow these to mix with the $Y_a$, via
one-loop and higher-loop nondiagonal propagator corrections, so that the
actual vector boson mass eigenstates would be, for $N_{g \ell}=1$,
\beqs
V_{a,heavy} & = & \cos\omega \ Y_a + \sin\omega \ V^{\hat \tau}_a \cr\cr
V_{a,light} & = & -\sin\omega \ Y_a +\cos\omega \ V^{\hat \tau}_a
\label{vymix}
\eeqs
and similarly in the case where $N_{g \ell} \ge 2$.  Since this mixing would be
forbidden at energy scales where $G_{SC}$ or $G_{SCC}$ is still an exact 
symmetry, and since the maximum of the relevant breaking scales is of order 
the highest ETC scale, $\Lambda_{ETC,max} \simeq 10^6$ GeV, it follows that 
\beq
|\omega| \simeq \frac{\Lambda_{ETC,max}}{M_{GU}} \ll 1 \ . 
\label{omega}
\eeq
Hence, the mixing would lead to a diagram for nucleon decay with a propagator
for $V_{a,light}$, $\simeq 1/\Lambda_{ETC,max}^2$ multiplied by the mixing
factors for each of the two vertices involving SM fermions, which have size
$\lsim (\Lambda_{ETC,max}/M_{GU})^2$.  The product is then $\lsim 1/M_{GU}^2$,
the same as for the usual contribution from $V_{a,heavy}$.  Hence (given that
$M_{GU}$ is sufficiently large so that the usual contributions to nucleon decay
are not excessive) this mixing does not significantly increase the rate of
nucleon decay.  This shows that these ETC gauge bosons can, indeed, have
ETC-scale masses, as required to give SM fermions their masses.

  We next consider the constraint that there be no anomaly in gauged currents
of the unified theory invariant under the group $G$. For this purpose, we
define a $(N-1)$-dimensional vector of anomalies
\beq
{\bf a} = (A([1]_N),...,A([N-1]_N))
\label{adef}
\eeq
where $A([k]_N)$ is given in eq. (\ref{anomaly}) in the appendix. 
Then the constraint that there be no $G$ gauge anomaly is the condition
\beq 
{\bf n} \cdot {\bf a} = 0 \ .
\label{n.a}
\eeq
which is a diophantine equation for the components of the vector of
multiplicities ${\bf n}$, subject to the constraint that the components $n_k$
are non-negative integers.  Geometrically, if ${\bf a}$ and ${\bf n}$ were
vectors in ${\mathbb R}^{N-1}$, then the solution set of eq. (\ref{n.a}) would
be the $(N-2)$-dimensional subspace of ${\mathbb R}^{N-1}$ orthogonal to the
vector ${\bf a}$; the situation here is more complicated because of the
diophantine requirement that $n_k$ is a nonnegative integer. The actual
solution is also subject to additional conditions, as we shall discuss shortly.

The most natural way to satisfy the third requirement, that the TC-singlet,
SM-nonsinglet fermions should form a well-defined set of SM generations, is to
impose this separately on the subset of these fermions that arise from the
fermion representations of $G$ and on the complementary subset that arise from
the sequential symmetry breaking of SU($N_{SCC})$. The fermion representations
of $G$ transform according to $({\cal R}_{SC},{\cal R}_{GU})$ with respect to
the subgroup decomposition (\ref{scxgut}).  In terms of these, the condition on
the former subset yields the two conditions
\beq 
N_{(1,\bar 5)} = N_{(1,10)}
\label{5bareq10}
\eeq
and, for nonsinglet ${\cal R}_{SU(5)_{GU}}$, 
\beq
N_{(1,{\cal R}_{SU(5)_{GU}})} = 0 \ \  {\rm if} \ \ {\cal R}_{SU(5)_{GU}}
 \ne \bar 5 \ \ {\rm or} \ \ 10 \ , 
\label{nrzero}
\eeq
i.e., the number of TC-singlet left-handed (anti)fermions from these
representations transforming as $\bar 5$ and $10$ of SU(5)$_{GU}$ must be equal
and the theory must not contain any other TC-singlet, SM-nonsinglet fermions.
Regarding the fermions in the complementary subset, we note that the breaking
of SU($N_{SCC}$) can be viewed as the breaking of the SU($N_{SC}$) part of this
group, since the SU(3)$_c$ part remains unbroken.  For these, the 
requirements that we impose are the analogues of eqs. (\ref{5bareq10}) and 
(\ref{nrzero}) with the multiplets considered to refer to 
$({\cal R}_{TC},{\cal R}_{GU})$. 

As noted earlier, in models of type (3), for a given choice of fermion
representations of the unified group $G$, it is not guaranteed that the
resultant (TC-singlet) SM fermions come in well-defined generations, and even
for a choice which does satisfy this constraint, it is not guaranteed that the
model can accomodate three such generations.  In both respects, these models
are different from models of type (1) and (2), where one can automatically
satisfy both of these conditions.

We now incorporate the constraint that the standard-model fermions that arise
from the fermion representations of $G$ comprise well-defined generations.
Each of these SM fermion generations is equivalent to the set $\{(1,\bar 5) +
(1,10)\}$ of representations of left-handed fermions, under the direct product
${\rm SU}(N_{SC}) \times {\rm SU}(5)_{GU}$. A $(1,\bar 5)$ representation
arises in two ways (i) from a $[N_{SC}+4]_N = [N-1]_N \approx [\bar 1]_N$
representation, $\psi_{i,L}$, when $i$ takes values in SU(5)$_{GU}$; and (ii)
from a $[4]_N$ representation, $\psi^{i_1 i_2 i_3 i_4}_L$, when all of the
indices take on values in SU(5)$_{GU}$.  If one were to choose $N_{SC}=0$ and
hence $N=5$, these sources would coincide; for the relevant case of a
nonabelian TC group, for which $N_{SC} \ge N_{TC} \ge 2$, they constitute two
different sources.  Hence, for $N_{SC} \ge 2$,
\beq
N_{(1,\bar 5)} = n_{_{N_{SC}+4}}+n_4 
\label{n15bar}
\eeq
where, equivalently, $n_{_{N_{SC}+4}} = n_{_{N-1}}$. 
A (1,10) representation also arises in two ways: (i) from a $[2]_N$
representation, $\psi^{i_1 i_2}_L$, when both of the indices take on values in
SU(5)$_{GU}$, and (ii) from a $[N_{SC}+2]_N$ representation, when $N_{SC}$ of
the indices take on values in SU($N_{SC}$), thereby producing a singlet under
this group, and the remaining two indices take on values in SU(5)$_{GU}$.
(Since $[N_{SC}+2]_N \approx [\overline{N-3}]_N$, one can equivalently describe
source (ii) as arising from $\psi_{i_1 i_2 i_3,L}$ when all of the three
indices take on values in SU(5)$_{GU}$.) Again, for $N_{SC}=0$ and hence $N=5$,
these sources (i) and (ii) coincide; for the relevant nonabelian case $N_{TC}
\ge 2$ and hence $N_{SC} \ge 2$, they are different, so that
\beq
N_{(1,10)} = n_2 + n_{_{N_{SC}+2}} \ . 
\label{n10}
\eeq
Thus, the requirement that the left-handed SC-singlet, SM-nonsinglet
(anti)fermions comprise equal numbers of $(1,\bar 5)$ and (1,10)'s implies the
condition 
\beq
n_{_{N_{SC}+4}} + n_4 = n_2 + n_{_{N_{SC}+2}}
\label{generationcondition}
\eeq
and the number of SM fermion generations $N_{gh}$ produced by the 
representations of $G$ is given by either side of this equation; 
\beq
N_{gh} = n_2 + n_{_{N_{SC}+2}} \ . 
\label{ngenh}
\eeq
The remaining $N_{g \ell}$ generations of SM fermions arise via the breaking of
$G_{SCC}$ and/or $G_{SC}$. 

We next determine the implications of the constraint (\ref{nrzero}) excluding
SC-singlet fermions that have unphysical nonsinglet SM transformation
properties.  These can be identified in terms of their SU(5)$_{GU}$
representations. Since the only possibilities for nonsinglet $[k]_5$ are
$[1]_5=5$, $[2]_5=10$, $[3]_5=\overline{10}$, and $[4]=\bar 5$, these
unphysical representations are the $(1,5)$ and $(1,\overline{10})$.  Now a
$[1]_N$ representation, $\psi^i_L$, yields a (1,5) when the index $i$ takes on
values in SU(5)$_{GU}$.  Further, a $[N_{SC}+1]_N$ representation,
$\psi^{i_1...i_{N_{SC}+1}}_L$, also yields a (1,5) when $N_{SC}$ of the indices
take on values in SU($N_{SC}$), thereby yielding an SC-singlet, and the one
remaining index takes on values in SU(5)$_{GU}$.  The requirement that there be
no (1,5)'s is therefore
\beq
n_1 = 0, \quad n_{_{N_{SC}+1}}=0 \ . 
\label{no5}
\eeq
A $[3]_N$ representation, $\psi^{i_1 i_2 i_3}_L$, yields a $(1,\overline{10})$
if all of the three indices take values in SU(5)$_{GU}$.  In addition, a
$[N_{SC}+3]_N$ representation, $\psi^{i_1...i_{N_{SC}+3}}_L$, yields a
$(1,\overline{10})$ in the case when $N_{SC}$ of the indices take on values in
SU($N_{SC}$) and the remaining three indices take on values in SU(5)$_{GU}$. 
(Since $[N_{SC}+3]_N \approx [\overline{N-2}]_N$, the latter source is 
equivalent to $\psi_{i_1 i_2,L}$ with both indices taking on values in
SU(5)$_{GU}$.) Hence, the requirement that there be no $(1,\overline{10})$ is
\beq
n_3 = 0, \quad n_{_{N_{SC}+3}}=0 \ . 
\label{no10bar}
\eeq

The representations $[1]_N$ and $[N-1]_N \approx [\bar 1]_N$, when decomposed
with respect to the subgroup (\ref{wxscc}), will yield a term (2,1), i.e., a
doublet under SU(2)$_w$ which is a singlet under SU($N_{SCC}$).  This is a
lepton doublet, such as ${\nu_e \choose e}_L$. The fact that it is a singlet
under SU($N_{SCC}$) means that neither of the component fermions couples
directly to the ETC gauge bosons, and hence both have strongly suppressed
masses.  This is acceptable for the neutrino, but $m_e$ is only about a factor
of 10 less than $m_u$, so this strong mass suppression may be problematic for
the electron. In order to prevent this, one could require that $n_1 = 0$ and
$n_{N-1} = 0$.  We shall not do this here, but it should be borne in mind that
models with nonzero values for $n_1$ and/or $n_{N-1}$ will have this property.

In the present type-3 models, SC-singlet, SM-singlet fermions (1,1), which can
be identified as electroweak-singlet neutrinos, arise, in general, from two
sources: (i) $[N_{SC}]_N$, when all of the $N_{SC}$ indices take values in
SU($N_{SC}$); and (ii) $[5]_N$, when all of the indices take values in
SU(5)$_{GU}$.  In the special case $N_{SC}=5$, these each contribute. Hence,
\beq
N_{(1,1)}=n_{_{N_{SC}}}+n_5 \ . 
\label{n11}
\eeq
If $G_{SC}$ is the same as the TC group, then this is the full set of
TC-singlet, electroweak-singlet neutrinos, so that the the right-hand side of
eq. (\ref{n11}) should be nonzero.  If $N_{SC} > N_{TC}$, then
electroweak-singlet neutrinos can also arise from SC-nonsinglet representations
when the SC group breaks to the TC group.

The constraint concerning the breaking of ${\rm SU}(N_{SCC})$ and the behavior
of the SU($N_{TC})$ technicolor group that is operative below the lowest ETC
breaking scale entails several parts.  Since the TC theory emerges from the
breaking of the SCC theory, and since at the unification scale the squared
coupling $\alpha_{SCC} = \alpha$ is small, one wants the SCC theory to be
asymptotically free in order for $\alpha_{SCC}$ to increase as the energy scale
decreases, yielding, after breakings, a TC coupling that is sufficiently large
to produce the eventual technifermion condensate.  The asymptotic freedom of
the SCC theory is required if, as assumed here, one or more of the sequential
breakings of the SU($N_{SCC}$) theory are self-breakings.  The constraint
$\beta_{TC} < \beta_{SU(3)_c} < 0$ in the original approach to the unification
of TC and SM gauge symmetries does not appear here because the SU(3)$_c$ group
is subsumed within the SU($N_{SCC}$) group in the relevant range of energies
$\Lambda_{ETC,max} < \mu < M_{GU}$.  Note that, since $N_{SCC} \ge 5$, it
follows that $b_0^{(SCC)} - b_0^{(2)} = N_{SCC}-2 \ge 3$, where here
$b_0^{(2)}$ is the leading coefficient of the only other nonabelian subgroup of
$G$, namely SU(2)$_w$ (cf. eq. (\ref{wxscc})).  Hence, provided that the
SU($N_{SCC}$) theory is asymptotically free, its beta function is more negative
than that of the SU(2)$_w$ sector, as should be the case to account for the
observed values of the SM gauge couplings at the electroweak scale.

Before analyzing specific models, we mention some challenges that can be
anticipated at the outset.  First, for models with $N_{gh}=3$ so that
$N_{SC}=N_{TC}$, there is a single effective ETC mass scale that governs the
origin of the SM fermion masses, namely the scale at which the breaking 
\beq
{\rm SU}(N_{SCC}) \to G_{TC} \times {\rm SU}(3)_c
\label{gsccbreak}
\eeq
occurs.  (There would, in general, also be a U(1) factor; here we concentrate
on the nonabelian symmetries.)  The only other scale that enters into the
generation of the masses for these SM fermions is the technicolor scale.  With
only a single effective ETC scale to work with, one cannot satisfactorily
reproduce the observed SM fermion mass hierarchy.  In models with $N_{g \ell }
\ge 1$, there is, {\it a priori}, the formal possibility of having enough ETC
mass scales to produce the observed SM generational hierarchy, making use of
the sequential breaking scales of $G_{SCC}$ and/or $G_{SC}$ that are supposed
to yield the $N_{g \ell}$ additional generations.

However, when we actually examine these models based on a simple unification
group $G$ (without auxiliary groups such as hypercolor), we find that the
requisite sequential dynamical symmetry breaking of ${\rm SU}(N_{SCC})$ is
unlikely to occur. The breaking of $G_{SCC}$ should eventually yield, after the
sequential self-breaking, the residual exact nonabelian symmetry group on the
right-hand side of eq. (\ref{gsccbreak}).  Given that the ${\rm SU}(N_{SCC})$
theory is asymptotically free, the associated gauge coupling $\alpha_{SCC}$
will increase as the energy scale decreases from the unification scale,
$M_{GU}$, and, when $\alpha_{SCC}$ is sufficiently large, the theory will form
bilinear fermion condensate(s).  If the ${\rm SU}(N_{SCC})$ gauge interaction
is vectorlike, i.e., if (neglecting other gauge interactions) the nonsinglet
fermion content of ${\rm SU}(N_{SCC})$ consists of the set of left-handed
fermions $\{ \sum_{{\cal R}} {\cal R} + \bar {\cal R} \}$, then the most
attractive channel for this condensation process is
\beq
{\cal R} \times \bar {\cal R} \to 1 \ , 
\label{rrbar1}
\eeq
i.e., it yields a condensate that is a singlet under ${\rm SU}(N_{SCC})$.
Hence, the model would fail to break ${\rm SU}(N_{SCC})$ at all, let alone to
the residual subgroup (\ref{gsccbreak}).  With the same initial set of
representations, one also has the channel
\beq
{\cal R} \times \bar {\cal R} \to Adj \ , 
\label{rrbaradj}
\eeq
where here $Adj$ refers to the adjoint representation of SU($N_{SCC}$).  This
channel could lead to the desired breaking of SU($N_{SCC}$) in
eq. (\ref{gsccbreak}).  However, channel (\ref{rrbar1}) is always more
attractive than channel (\ref{rrbaradj}). From our studies of specific models,
we find that in most cases where $G_{SCC}$ is asymptotically free, it is
vectorlike, and hence, applying the MAC criterion, one would conclude that the
necessary dynamical symmetry breaking would not take place.  Even in a model
with $N_{SC}=5$ and with ${\bf n}$ given in eq. (\ref{nsu10}), where
SU($N_{SCC}$) is an asymptotically free chiral gauge theory, we find that it is
unlikely to break in the desired manner.

A related problem with the dynamical symmetry breaking is that in many cases,
not only does the condensation in the most attractive channel not break
SU($N_{SCC}$), it breaks SU(2)$_w$ at a scale which is higher than the ETC
scales where SU($N_{SCC}$) should break.  A possible way to avoid undesired
condensation channels of this sort could be to invoke a ``generalized most
attractive channel'' (GMAC) criterion \cite{at94,nt,ckm}, which makes use of
vacuum alignment and related energy minimization arguments to suggest that if
the condensate formation can avoid breaking a certain symmetry, it will
\cite{tcvac}.

Yet another complication can occur in cases where $N_{SC}$ is odd so that
$N_{SCC}$ is even, say $N_{SCC}=2p$.  In these cases, there can occur a most
attractive channel of the form
\beq
[p]_{2p} \times [p]_{2p} \to 1 
\label{ppchannel}
\eeq
with
\beq
\Delta C_2 = 2C_2([p]_{2p})= \frac{p(2p+1)}{2} = \frac{N_{SCC}(N_{SCC}+1)}{4} \
. 
\label{deltac2ppchannel}
\eeq
The associated condensate is
\beq
\langle \epsilon_{i_1...i_{2p}} \psi^{1_1...i_p \ T}_L C 
\psi^{1_{p+1}...i_{2p}}_L \rangle \ . 
\label{ppcondensate}
\eeq
This condensate is symmetric (antisymmetric) under interchange of
$\psi^{1_1...i_p}_L$ and $\psi^{1_{p+1}...i_{2p}}_L$ if $p$ is even (odd).
Since the condensate (\ref{ppcondensate}) is invariant under $G_{SCC}$, it is,
{\it a fortiori}, invariant under SU($N_{SC}$) and SU(3)$_c$.  The only way to
construct the requisite SU(3)$_c$-invariant contractions involves product(s)
$\epsilon_{\ell m n}d^\ell d^m d^n$, each of which has weak hypercharge $Y=-2$
(and electric charge $-1$).  Hence, this condensate violates weak hypercharge
and electric charge.  It may be noted that in a hypothetical world in which
only the SU(2)$_w$ interaction were strongly coupled, the same kind of
violation would presumably occur. Consider, say, the first two generations of
lepton doublets, $\psi_{g1,L} = {\nu_e \choose e}$ and $\psi_{g2,L} = {\nu_\mu
\choose \mu}$.  These would form an SU(2)$_w$-invariant condensate of type
(\ref{ppcondensate}) with $p=1$, namely
\beq
\langle \epsilon_{jk} \psi^{j \ T}_{1g,L} C \psi^k_{2g,L} \rangle = 
2\langle \nu_{eL}^T C \mu_L - e^T_L C \nu_{\mu L} \rangle \ . 
\label{weakcondensate}
\eeq
where $j,k$ are SU(2)$_w$ indices. 

\subsection{$N_{SC}=2$, $G={\rm SU}(7)$}

 We proceed to study a number of specific models of type (3) to explore their
properties.  We begin by considering the minimal nontrivial case, $N_{SC}=2$.
Since this is the smallest value for a nonabelian group, it follows
that $G_{SC} = G_{TC}$ and hence we denote $G_{SCC} \equiv G_{TCC}$; further,
it follows that $N_{g \ell}=0$, so that an acceptable model would have to have
$N_{gh}=3$.  The vector ${\bf n}$ has the form ${\bf n} = (n_1,...,n_6)_7$.
From eqs. (\ref{no5}) and (\ref{no10bar}) we have $n_1=n_3=n_5=0$.  Equation
(\ref{generationcondition}) yields $n_6=n_2$.  The no-anomaly condition,
eq. (\ref{n.a}), reads $3n_2-2n_4-n_6=0$; substituting $n_6=n_4$ in this
equation gives the result $n_2=n_4$.  Hence, ${\bf n} = n_2(0,1,0,1,0,1)_7$
Taking $n_2=1$ yields
\beq
{\bf n} = (0,1,0,1,0,1)_7 \ . 
\label{nsu7}
\eeq
More generally, for an SU($N$) group with $N$ odd, say $N=2n+1$, the chiral
fermion content $\{ f \} = \sum_{\ell=1}^n \ [2\ell]_N$ is anomaly-free
\cite{g79}.  This property was used in Ref. \cite{fs} for a study of the
possible unification of TC and SM symmetries in SU(7) and SU(9). For the
fermion set in eq. (\ref{nsu7}) the number of SM generations is given by
eq. (\ref{ngenh}) as $N_{gen.}=2n_2=2$, so the requirement that $N_{gen.}=3$
cannot be satisfied, and this model is not acceptable.  One could,
nevertheless, consider it as a toy model.  The simplest special case of this
toy model has $n_2=1$, so that there would be two SM generations.  Taking the
next higher value, $n_2=2$ is not acceptable because it would yield the
unphysical result of four SM generations.  Applying eq. (\ref{n11}), we note
that there is one electroweak-singlet neutrino.

With respect to the group (\ref{scxgut}), namely, 
\beq
{\rm SU}(2)_{TC} \times {\rm SU}(5)_{GU} \ , 
\label{scxgutsu7}
\eeq
the fermions have the following decompositions:
\beq
[2]_7 = (1,1) + (2,5) + (1,10)
\label{2_su7}
\eeq
\beq
[4]_7 \approx [\bar 3]_7 = (1,\bar 5) + (2,\overline{10}) + (1,10)
\label{4_su7}
\eeq
\beq
[6]_7 \approx [\bar 1]_7 = (2,1) + (1,\bar 5)
\label{6_su7}
\eeq
where here and below we use the equivalences $2 \approx \bar 2$ for SU(2) and
(\ref{nkn}) for SU($N$).  

The technifermions in this model are 
\beqs
& & U^a_{\tau L} \ , \quad D^{\tau a}_L \ , \quad
     { U^{\tau a} \choose D^{\tau a}}_R  \cr\cr
& & N_{\tau L} \ , \quad E_{\tau L} \ , \quad
 {N_\tau \choose E_\tau}_R
\label{tcfermionssu7}
\eeqs
where, as before, $\tau=1,2$ is the technicolor index and $a=1,2,3$ is the
color index.  Note how, in accordance with our general discussion above, the
left- and right-handed chiral components of the charge 2/3 techniquark, $U_L$
and $U_R$, transform according to relatively conjugate representations of
SU(2)$_{TC}$.  In this case, since SU(2) has only (pseudo)real representations,
these are equivalent, and the technicolor theory is a vectorial gauge theory.

Since both the subgroup (\ref{scxgut}) and the subgroup (\ref{wxscc}), 
\beq
{\rm SU}(2)_w \times {\rm SU}(5)_{TCC} \ , 
\label{wxsccsu7}
\eeq
are abstractly SU(2) $\times$ SU(5), the fermions have formally the same
decompositions with respect to (\ref{wxsccsu7}) as in eqs.
(\ref{2_su7})-(\ref{6_su7}), although the component fields are different.  It
should be noted that a fermion that is a singlet under color and technicolor,
and hence is a lepton, can occur as a component of a TCC nonsinglet
representation.  For example, with respect to the subgroup (\ref{wxsccsu7}),
the $[2]_7$ yields a term 
\beq
(1,10) = \left( \begin{array}{ccccc}
      0  & \nu^c   & D^{11} & D^{12} & D^{13}  \\
         & 0       & D^{21} & D^{22} & D^{23}  \\
         &         &  0     & u^c_3  & -u^c_2  \\
         &         &        & 0      & u^c_1   \\
         &         &        &        & 0 \end{array} \right )_L
\label{2bar10a}
\eeq
where the upper indices $\tau a$ on $D^{\tau a}$ refer to technicolor and
color, and the entries left blank are equal to minus the transposed
entries. The $\nu^c_L$ field illustrates the general point made above.

The SU(5)$_{TCC}$ interaction (neglecting other interactions) is vectorial,
consisting of the left-handed fermion content $2\{ 5 + \bar 5 + 10 +
\overline{10} \}$.  The SU(5)$_{TCC}$ gauge interaction is asymptotically free,
with leading beta function coefficient $b^{(TCC)}_0 = 13$.  Hence, as the
energy scale decreases below $M_{GU}$, $\alpha_{TCC}$ increases to the point
where the theory forms bilinear fermion condensates. Let us consider
condensations of these fermions, with the representations classified according
to the subgroup of eq. (\ref{wxsccsu7}) as $({\cal R}_{SU(2)_w},{\cal
R}_{SU(5)_{TCC}})$.  In this notation, the most attractive channel (MAC) is
\beq
(1,10) \times (2,\overline{10}) \to (2,1) \ . 
\label{su5channel1}
\eeq
If, indeed, this condensate formed, it would rule out this SU(7) model, since
it would break SU(2)$_w$, at much too high a scale; indeed, this
SU(2)$_w$-breaking scale would be greater than the ETC scales where
SU(5)$_{TCC}$ should break, clearly an unphysical situation.  The
attractiveness of the channel, as measured by $\Delta C_2$, is given by $\Delta
C_2 = 36/5$.  If one uses as a rough guide the critical value of $\alpha_{TCC}$
given by the Schwinger-Dyson gap equation (eq. (\ref{alfcrit}) in the
appendix), one finds the value $\alpha_{TCC}=5\pi/54 \simeq 0.3$.  As an
illustrative value, assume that $\alpha_{TCC}=\alpha_{GU} \simeq 0.04$ at
$M_{GU}$.  Substituting the above critical value into eq. (\ref{mucrit}), one
obtains the rough estimate that $\alpha_{TCC}$ increases to the critical value
for the condensate (\ref{su5channel1}) to form as the scale $\mu$ decreases
through the value $\mu_c \simeq (3 \times 10^{-5})M_{GU}$.  For the
hypothetical unification scale $M_{GU} = 10^{16}$ GeV, this would mean that
electroweak symmetry would be broken at $\sim 10^{11}$ GeV, clearly far too
high a scale.  In addition, the channel (\ref{su5channel1}) would fail to break
the SU(5)$_{SCC}$ group to ${\rm SU}(2)_{TC} \times {\rm SU}(3)_c$.

Other condensation channels which are, {\it a priori} possible, are listed
below, together with their $\Delta C_2$ values,
\beq
(1,10) \times (1,10) \to (1,\bar 5) \ , \quad \Delta C_2 = 24/5
\label{su5channel2}
\eeq
\beq
(1,\bar 5) \times (2,5) \to (2,1) \ , \quad \Delta C_2 = 24/5
\label{su5channel3}
\eeq
\beq
(1,\bar 5) \times (1,10) \to (1,5) \ , \quad \Delta C_2 = 18/5
\label{su5channel4}
\eeq
\beq
[(2,5) \times (2,\overline{10})]_a \to (1,\bar 5) \ , \quad \Delta C_2=18/5 
\label{su5channel5}
\eeq
\beq
[(2,5) \times (2,\overline{10})]_s \to (3,\bar 5) \ , \quad \Delta C_2=18/5 
\label{su5channel6}
\eeq
where the subscripts $a$ and $s$ in eqs. (\ref{su5channel5}) and
(\ref{su5channel6}) refer to antisymmetric and symmetric combinations of
representations.  None of these channels is acceptable in a viable model.
Consider, for example, channel (\ref{su5channel2}).  The condensate for this
channel is 
\beq
\langle \epsilon_{ijk \ell n} \psi^{jk \ T}_L C \psi^{\ell n}_L \rangle
\label{su5channel2condensate}
\eeq
where the indices are in SU(5)$_{TCC}$ (with the ordering as in
eq. (\ref{5genR})) and $\psi^{jk}_L$ is the fermion field transforming as
(1,10) under ${\rm SU}(2)_w \times {\rm SU}(5)_{TCC}$.  The free index must
take on one of the two SU(2)$_{TC}$ values, $i=1$ or $i=2$, in order to avoid
breaking SU(3)$_c$; with no loss of generality, we may choose $i=1$.  The
condensate (\ref{su5channel2condensate}) is then proportional to
\beq
\langle D^{21 \ T}_L C u^c_{1L} + D^{22 \ T}_L C u^c_{2L} + 
D^{23 \ T}_L C u^c_{3L} \rangle
\label{su5channel2explicit}
\eeq
where the indices on $D^{\tau a}$ are as in eq. (\ref{2bar10a}).  This
condensate violates weak hypercharge and electric charge, and leaves only one
unbroken technicolor index, so that technicolor becomes an abelian
symmetry. Channel (\ref{su5channel3}) breaks SU(2)$_w$ at too high a scale and
fails to break SU(5)$_{TCC}$.  Channels (\ref{su5channel4}) and
(\ref{su5channel5}) leave technicolor as an abelian symmetry. Channel
(\ref{su5channel6}) breaks SU(2)$_w$ in the wrong way and at too high a scale,
and leaves technicolor as an abelian symmetry.

Thus, none of these condensation channels produces the desired symmetry
breaking of SU(5)$_{TCC}$ to ${\rm SU}(2)_{TC} \times {\rm SU}(3)_c$.  In the
absence of an actual mechanism to produce this breaking, it is difficult to
analyze the properties of the hypothetical resultant SU(2)$_{TC}$ theory.  With
the one SM family of technifermions listed in eq. (\ref{tcfermionssu7}), the
SU(2)$_{TC}$ interaction would be asymptotically free, with leading beta
function coefficient $b^{(TC)}_0 = 2$, but since this is smaller than the
corresponding $b^{(3)}_0$ for SU(3)$_c$ (which is $b^{(3)}_0=25/3 \simeq 8.3$
for this toy two-generation model), and $\alpha_{TC} = \alpha_{SU(3)_c}$ at the
energy scale where SU(5)$_{TCC}$ splits to ${\rm SU}(2)_{TC} \times {\rm
SU}(3)_c$, the color coupling would grow considerably faster than the
technicolor coupling as the energy scale decreased, leading to the unphysical
prediction that $\Lambda_{QCD} > \Lambda_{TC}$.

\subsection{$N_{SC}=3$, $G={\rm SU}(8)$}

Next, we consider $N_{SC}=3$, $G={\rm SU}(8)$, so that ${\bf n} =
(n_1,...,n_7)_8$.  In this case, {\it a priori}, one has the two options
$N_{TC}=N_{SC}=3$ with $N_{g \ell}=0$, or $N_{TC}=2$ with $N_{g \ell}=1$.
In general, eqs. (\ref{no5}) and (\ref{no10bar}) yield
\beq
n_1=n_3=n_4=n_6=0
\label{nzerosu8}
\eeq
and eq. (\ref{generationcondition}) reads
\beq
N_{gh}=n_2+n_5=n_4+n_7 \ . 
\label{ngenhsu8}
\eeq
The no-anomaly condition is
\beq
4n_2-5n_5-n_7=0 \ . 
\label{anomalysu8}
\eeq
For a given value of $N_{gh}=3-N_{g \ell}$, these are three
nondegenerate linear equations for the three quantities $n_2$, $n_5$, and
$n_7$.  We display the formal solution, with the understanding that it is
physical only for positive nonnegative integer values of the $n_k$:
\beqs
n_2 & = & \frac{2N_{gh}}{3} \cr\cr
n_5 & = & \frac{ N_{gh}}{3} \cr\cr
n_7 & = & N_{gh} \ . 
\label{ngensolsu8}
\eeqs
In order for $n_2$ and $n_5$ to be nonnegative integers, $N_{gh} = 0$ mod
3; the value $N_{gh}=0$ is not permitted because this would require 
$N_{g \ell}=3$, but $N_{g \ell} \le N_{SC}-(N_{TC})_{min}=N_{SC}-2=1$.
Hence, the only possibility is 
\beq
N_{gh}=3, \quad N_{g \ell}=0  \ , 
\label{ngenh3}
\eeq
whence
\beq
n_2=2, \quad n_5=1, \quad n_7=3 \ , 
\label{nsolsu8}
\eeq
so that
\beq
{\bf n} = (0,2,0,0,1,0,3)_8 \ , 
\label{nsu8}
\eeq
i.e., the fermion content of the model is 
\beq
\{ f \} = 2[2]_8 + [5]_8 + 3[7]_8 \approx
 2[2]_8 + [\bar 3]_8 + 3[\bar 1]_8 \ . 
\label{su8sol}
\eeq
Further, for this case, 
\beq
G_{SC}=G_{TC}={\rm SU}(3)_{TC}
\label{gscsu8}
\eeq
and 
\beq
G_{SCC}={\rm SU}(6)_{SCC} = {\rm SU}(6)_{TCC} \ , 
\label{gsccsu8}
\eeq
where, since $G_{SC}$ is just the technicolor group, we have indicated this
explicitly in the subscript. Note that, by eq. (\ref{n11}),
\beq
N_{(1,1)}=1 \ . 
\label{n11su8}
\eeq
In Table \ref{properties} we list properties of this model and others that we
have studied.

Let us analyze this SU(8) model further.  With respect to the subgroup
given by (\ref{scxgut}), viz., 
\beq
{\rm SU}(8) \supset {\rm SU}(3)_{TC} \times {\rm SU}(5)_{GU} \ , 
\label{scxgutsu8}
\eeq
we have the decomposition
\beq
[2]_8 = (\bar 3,1) + (3,5) + (1,10)
\label{2_8}
\eeq
\beq 
[5]_8 \approx [\bar 3]_8 = (1,1) + (3,\bar 5) + (\bar 3, \overline{10}) +
(1,10)
\label{5_8}
\eeq
\beq
[7]_8 \approx [\bar 1]_8 = (\bar 3,1) + (1,\bar 5) \ . 
\label{7_8}
\eeq
Since both technicolor and color are described by SU(3) subgroups of SU(8), the
theory is formally symmetric under the interchange of technicolor indices
$i=1,2,3$ and color indices, $i=4,5,6$, and eqs. (\ref{2_8})-(\ref{7_8}) also
describe the decomposition of the fermion representations with respect to
the subgroup ${\rm SU}(3)_c \times {\rm SU}(5)$, where this SU(5) involves
technicolor and electroweak indices.  The nonsinglet fermion content under
color or technicolor consists of 15 copies of $\{ 3 + \bar 3 \}$.  Evidently,
both color and technicolor are vectorial gauge symmetries.

With respect to the subgroup given by (\ref{wxscc}), namely
\beq
{\rm SU}(8) \supset {\rm SU}(2)_w \times {\rm SU}(6)_{TCC}
\label{wxsccsu8}
\eeq
the fermion representations have the decompositions 
\beq
[2]_8 = (1,1)+(2,[1]_6)+(1,[2]_6)
\label{2_8a}
\eeq
\beq
[5]_8 \approx [\bar 3]_8 =(1,[\bar 1]_6)+(2,[\bar 2]_6)+(1,[\bar 3]_6)
\label{5_8a}
\eeq
\beq
[7]_8 \approx [\bar 1]_8 = (2,1) + (1,[\bar 1]_6) \ . 
\label{7_8a}
\eeq
Here and below, it is convenient to use the $[k]_N$ notation for larger Lie
groups; the corresponding dimensionalities for representations of SU(6)$_{TCC}$
are $[2]_6=15$ and $[\bar 3]_6=\overline{20}$.  Thus, the
SU(6)$_{TCC}$ theory is vectorial, with nonsinglet fermion content (neglecting
other interactions) consisting of the set of (left-handed) fermions 
\beq
4([1]_6 + [\bar 1]_6)+2([2]_6 + [\bar 2]_6 + [\bar 3]_6) 
\label{su6scc}
\eeq
(Here, with respect to SU(6), $[3]_6 \approx [\bar 3]_6$.)  Let us assume that
SU(8) breaks in such a manner as to yield an effective theory at lower energies
that has a ${\rm SU}(2)_w \times {\rm SU}(6)_{TCC}$ symmetry (ignoring an
abelian factor).  Since the TCC theory is asymptotically free, with leading
beta function coefficient $b^{(TCC)}_0 = 12$, the TCC coupling increases as the
energy scale decreases.  Because the SU(6)$_{TCC}$ theory is vectorial, when
the energy scale decreases sufficiently that $\alpha_{TCC} \sim O(1)$, the TCC
interaction will naturally form SU(6)$_{TCC}$-invariant fermion condensates
rather than breaking to ${\rm SU}(3)_c \times {\rm SU}(3)_{TC}$, as is
necessary in order to separate color and the TC interaction.  The most
attractive channel, written in terms of representations of ${\rm SU}(2)_w
\times {\rm SU}(6)_{TCC}$, together with its $\Delta C_2$ value, is
\beq
(1,[\bar 3]_6) \times (1,[\bar 3]_6) \to (1,1) \ , \quad 
\Delta C_2 = \frac{21}{2} \ . 
\label{su6channel1}
\eeq
This condensation channel is of the form of (the conjugate of)
(\ref{ppchannel}) with $p=3$, $N_{SC}=6$.  By our general argument given above,
the associated condensate violates weak hypercharge and electric charge. If it
did occur, this, by itself, would rule out the present SU(8) model.  The rough
estimate for the corresponding critical coupling is $\alpha_{TCC,c} \simeq
2\pi/63 \simeq 0.1$ from the Schwinger-Dyson equation.  Substituting this into
eq. (\ref{mucrit}) with $\alpha_{TCC}=\alpha$ at $\mu=M_{GU}$ and using the
illustrative value for the unified coupling $\alpha=0.04$, we find that
$\alpha_{TCC}$ would be large enough for this condensate to form at $\mu_c
\simeq (3 \times 10^{-5})M_{GU}$, i.e., about $3 \times 10^{11}$ GeV for the
hypothetical unification scale $M_{GU}=10^{16}$.

Other possible condensation channels have smaller values of the attractiveness
measure $\Delta C_2$; they include the following, in order of descending 
$\Delta C_2$: 
\beq
(1,[2]_6) \times (2,[\bar 2]_6) \to (2,1) \ , \quad \Delta C_2 = 
\frac{28}{3} \simeq 9.3
\label{su6channel2}
\eeq
\beq
(1,[2]_6) \times (1,[\bar 3]_6) \to (1,[\bar 1]_6) \ , \quad \Delta C_2 = 7
\label{su6channel3}
\eeq
\beq
(2,[\bar 2]_6) \times (1,[\bar 3]_6) \to (2,[\bar 5]_6) \approx (2,[1]_6) \ , 
\Delta C_2 = 7
\label{su6channel4}
\eeq
\beq
(1,[\bar 1]_6) \times (1,[2]_6) \to (1,[1]_6) \ , \quad \Delta C_2 =
\frac{35}{6} \simeq 5.8 
\label{su6channel5}
\eeq
\beq
(2,[1]_6) \times (1,[\bar 1]_6) \to (2,1) \ , \quad \Delta C_2 = \frac{35}{6} 
\ . 
\label{su6channel6}
\eeq
Channels (\ref{su6channel2}) and (\ref{su6channel6}) fail to break
SU(6)$_{TCC}$ and break SU(2)$_w$ at a scale higher than the ETC scale where
SU(6)$_{TCC}$ should break.  Channels (\ref{su6channel3}) and
(\ref{su6channel5}) break SU(6)$_{TCC}$ to SU(5)$_{TCC}$ rather than ${\rm
SU}(3)_c \times {\rm SU}(3)_{TC}$.  Channel (\ref{su6channel4}) breaks
SU(2)$_w$ at too high a scale and breaks SU(6)$_{TCC}$ to SU(5)$_{TCC}$. 

It may be noted that even if there were some way to produce a breaking of
SU(6)$_{TCC}$ that yielded a lower-energy theory invariant under ${\rm
SU}(3)_{TC} \times {\rm SU}(3)_c$, the interchange symmetry between of
SU(3)$_{TC}$ and SU(3)$_c$ would imply that the respective technicolor and
color gauge couplings would evolve in the same way as the energy decreases
below the scale at which this condensate occurred.  Both of these sectors are
asymptotically free, and condensates would form, but the model would still not
be realistic, since the scale $\Lambda_{TC}$ would be the same as
$\Lambda_{QCD}$.

\subsection{$N_{SC}=4$, $G={\rm SU}(9)$}

Here we consider $N_{SC}=4$, so that $G={\rm SU}(9)$ and ${\bf n} =
(n_1,...,n_8)_9$.  {\it A priori}, one has three possibilities regarding the
origins of the SM fermion generations: (i) $N_{gh}=3$, or equivalently,
$N_{g \ell}=0$, whence $N_{TC}=4$; (ii) $N_{gh}=2$ so that
$N_{g \ell}=1$, which would be associated with a breaking of ${\rm
SU}(4)_{SC}$ to ${\rm SU}(3)_{TC}$; (iii) $N_{gh}=1$, or equivalently,
$N_{g \ell}=2$, so that one SM generation would arise initially from the
representations of $G$, and the other two would arise via the sequential
breaking ${\rm SU}(4)_{SC} \to {\rm SU}(3)_{SC}$ and then ${\rm SU}(3)_{SC} \to
{\rm SU}(2)_{TC}$.  In general, equations (\ref{no5}) and (\ref{no10bar}) yield
\beq
n_1=n_3=n_5=n_7=0
\label{nzerosu9}
\eeq
and eq. (\ref{generationcondition}) is 
\beq
N_{gh}=n_2+n_6=n_4+n_8 \ . 
\label{ngenhsu9}
\eeq
The condition of zero gauge anomaly (\ref{n.a}) is 
\beq
5(n_2+n_4)-9n_6-n_8=0 \ . 
\label{anomalysu9}
\eeq
For a given value of $N_{gh}=3-N_{g \ell}$, these are three
nondegenerate linear equations for the four quantities $n_2$, $n_4$, $n_6$, and
$n_8$.  Taking $n_2$, say, as the independent variable, initially free to take
on values $n_2=0,1,...,N_{gh}$, we find the formal solution of these
equations to be
\beqs
n_6 & = & N_{gh}-n_2 \cr\cr 
n_4 & = & \frac{1}{3}\left ( 5N_{gh}-7n_2 \right ) \cr\cr
n_8 & = & \frac{1}{3}\left (7n_2 - 2N_{gh} \right ) \ . 
\label{nsolsu9}
\eeqs
Consider first the {\it a priori} possible value $N_{gh}=3$, whence
$n_4=5-(7/3)n_2$ and $n_8=(7/3)n_2-2$.  In order for $n_2$ and $n_8$ to be
nonnegative integers, $n_2 = 0$ mod 3.  Now $n_2$ cannot be zero, because this
would make $n_8$ negative.  But $n_2$ also cannot be equal to 3, because this
would make $n_4$ negative.  Hence, the value $N_{gh}=3$ is not allowed.

Consider next the value
\beq
N_{gh}=2
\label{ngenh2su9}
\eeq
whence $N_{g \ell}=1$ and 
\beq
G_{SC}={\rm SU}(4)_{SC}, \quad G_{TC}={\rm SU}(3)_{TC} \ . 
\label{gsctc3}
\eeq
Equations (\ref{nsolsu9}) yield $n_4=(1/3)(10-7n_2)$ and $n_8=(1/3)(7n_2-4)$.
From its formal range for this case, $n_2=0,1,2$, the only allowed value is
$n_2=1$ which gives $n_4=n_8=n_6=1$, so that 
\beq
{\bf n} = (0,1,0,1,0,1,0,1)_9
\label{nsu9}
\eeq
i.e., the fermion content is given by $\{f\}=[2]_9+[4]_9+[6]_9+[8]_9$.  
For this case $N_{(1,1)}=1$.  

Finally, we examine the minimal possible value, $N_{gh}=1$, corresponding
to the maximal possible value of $N_{g \ell}$, namely
$N_{g \ell}=N_{SC}-(N_{TC})_{min}=4-2=2$.  Here eqs. (\ref{nsolsu9}) read
$n_4=(1/3)(5-7n_2)$ and $n_8=(1/3)(7n_2-2)$, where, {\it a priori}, $n_2$ can
take values in the set $\{0,1\}$ .  Evidently, neither of these values would
make $n_4$ and $n_8$ nonnegative integers and hence neither is allowed.

Let us return to the case with $N_{gh}=2$.  An initial comment is that the
corresponding value $N_{TC}=3$ is disfavored, relative to $N_{TC}=2$, since it
leads to larger technicolor contributions to precision electroweak quantities
and could reduce the likelihood of walking behavior for the technicolor theory.
Notwithstanding this concern, let us investigate this case.  The decomposition
of the fermion representations with respect to the subgroup (\ref{scxgut}),
which for this case reads 
\beq
{\rm SU}(4)_{SC} \times {\rm SU}(5)_{GU} \ , 
\label{scxgutsu9}
\eeq
is listed below: 
\beq
[2]_9 = (6,1) + (4,5) + (1,10)
\label{2_9}
\eeq
\beq
[4]_9 = (1,1) + (\bar 4,5) + (6,10) + (4, \overline{10}) + (1,\bar 5)
\label{4_9}
\eeq
\beq
[6]_9 \approx [\bar 3]_9 = (4,1)+(\bar 6,\bar 5)+(\bar 4,\overline{10})+(1,10)
\label{6_9}
\eeq
\beq
[8]_9 \approx [\bar 1]_9 = (\bar 4,1) + (1,\bar 5) \ . 
\label{8_9}
\eeq
(Note that $6 \approx \bar 6$, i.e., $[2]_4 \approx [\bar 2]_4$, in
SU(4)$_{SC}$.)  This decomposition shows that, neglecting the $G_{GU}$
couplings relative to those of SU(4)$_{SC}$, the latter interaction is
vectorial, involving nonsinglet fermions comprising a set of 16 copies of $\{4
+ \bar 4 + 6\}$ left-handed fermions.

As before, we consider the implications of a scenario in which the unified
group, here SU(9), breaks to yield a theory at lower energy scales that is
invariant under the gauge symmetry (ignoring an abelian factor) of the form
(\ref{wxscc}), which for the present model is
\beq
{\rm SU}(2)_w \times {\rm SU}(7)_{SCC} \ . 
\label{wxsccsu9}
\eeq
With respect to this direct product symmetry group, the fermions have the
decomposition
\beq
[2]_9 = (1,1) + (2,[1]_7) + (1,[2]_7)
\label{2_9a}
\eeq
\beq
[4]_9 = (1,[2]_7) + (2,[3]_7) + (1,[\bar 3]_7)
\label{4_9a}
\eeq
\beq
[6]_9 \approx [\bar 3]_9 = (1, [\bar 1]_7) + (2,[\bar 2]_7) + (1,[\bar 3]_7)
\label{6_9a}
\eeq
\beq
[8]_9 \approx [\bar 1]_9 = (2,1) + (1,[\bar 1]_7) \ . 
\label{8_9a}
\eeq
Here, the dimensionalities include ${\rm dim}([2]_7)=21$ and ${\rm
dim}([3]_7)=35$. Thus, the SU(7)$_{SCC}$ theory is vectorial, with nonsinglet 
fermion content (neglecting other interactions) given by
\beq
2 \{ [1]_7 +[\bar 1]_7 + [2]_7 + [\bar 2]_7 + [3]_7+[\bar 3]_7 \} \ . 
\label{su7scc_su9}
\eeq
The SU(7)$_{SCC}$ gauge interaction is asymptotically free, with leading beta
function coefficient $b^{(SCC)}_0 = 13/3$.  However, as in the models that we
studied above, because of the vectorlike nature of the SCC gauge symmetry, when
the energy decreases sufficiently so that $\alpha_{SCC}$ grows large enough to
produce fermion condensates, these will preferentially be in the channels
${\cal R} \times \bar {\cal R} \to 1$, where ${\cal R}$ refers to the
SU(7)$_{SCC}$ representation.  Thus, these preserve the SU(7)$_{SCC}$
invariance rather than breaking it down to a direct product which includes
${\rm SU}(3)_c \times {\rm SU}(4)_{SC}$, as is necessary to separate color from
the strongly coupled SU(4)$_{SC}$ group.  With respect to the ${\rm SU}(2)_w
\times {\rm SU}(7)_{SCC}$ subgroup, the most attractive channel, with its
measure of attractiveness $\Delta C_2$, is
\beq
(2,[3]_7) \times (1,[\bar 3]_7) \to (2,1) \ , \quad \Delta C_2 = \frac{96}{7}
\simeq 13.7 \ . 
\label{su7channel1}
\eeq
In addition to its failure to break the SU(7)$_{SCC}$ symmetry, this MAC breaks
SU(2)$_w$ at a scale that would be higher than the ETC scales where the
SU(7)$_{SCC}$ should break.  The same problems characterize a channel with a
somewhat smaller value of $\Delta C_2$, namely, 
\beq
(1,[2]_7) \times (2,[\bar 2]_7) \to (2,1) \ , \quad \Delta C_2 = \frac{80}{7}
\simeq 11.4 \ . 
\label{su7channel2}
\eeq

As before, one can examine other possible condensation channels with still
smaller values of $\Delta C_2$, which include
\beq
(1,[2]_7) \times (1,[\bar 3]_7) \to (1,[\bar 1]_7) \ , \quad \Delta C_2 =
\frac{74}{7} \simeq 10.6
\label{su7channel3}
\eeq
\beq
[(2,[2]_7) \times (2,[\bar 3]_7)]_a \to (1,[\bar 1]_7) \ , \quad \Delta C_2 =
\frac{74}{7}
\label{su7channel4}
\eeq
and 
\beq
[(2,[2]_7) \times (2,[\bar 3]_7)]_s \to (3,[\bar 1]_7) \ , \quad \Delta C_2 =
\frac{74}{7} \ . 
\label{su7channel5}
\eeq
Channel (\ref{su7channel5}) is unacceptable because it breaks SU(2)$_w$ in the
wrong way and at too high a scale.  Channels (\ref{su7channel3}) and
(\ref{su7channel4}) are not forbidden but would only break SU(7)$_{SCC}$ to
SU(6)$_{SCC}$, thereby necessitating a further breaking to ${\rm SU}(3)_c
\times {\rm SU}(3)_{TC}$; moreover, because of their subdominant $\Delta C_2$
values, it is difficult to make a convincing argument that they would
predominate.  Another conceivable breaking pattern is ${\rm SU}(7)_{SCC} \to
{\rm SU}(3)_c \times {\rm Sp}(4)$ \cite{fs}, but it is not clear what
dynamical fermion condensation channel could produce this breaking. 

\subsection{$N_{SC}=5$, $G={\rm SU}(10)$}

We proceed to examine the case where $N_{SC}=5$, so that $G={\rm SU}(10)$ and
${\bf n} = (n_1,...,n_9)_{10}$.  {\it A priori}, one has four possibilities for
the manner in which the SM fermion generation arise, as specified by
$(N_{gh},N_{g \ell},N_{TC})$, namely (i) (3,0,5), (ii) (2,1,4), (iii) (1,2,3),
and (iv) (0,3,2).  The minimization of technicolor contributions to electroweak
corrections favor the last of these options. The conditions (\ref{no5}) and
(\ref{no10bar}) forbidding $5_L$ and $\overline{10}_L$ yield
\beq
n_1=n_3=n_6=n_8=0
\label{nzerosu10}
\eeq
and eq. (\ref{generationcondition}) is 
\beq
N_{gh}=n_2+n_7=n_4+n_9 \ . 
\label{ngenhsu10}
\eeq
The condition of zero gauge anomaly (\ref{n.a}) is 
\beq
6n_2+14(n_4-n_7)-n_9=0 \ . 
\label{anomalysu10}
\eeq
For a given value of $N_{gh}=3-N_{g \ell}$, these are three nondegenerate
linear equations for the five quantities $n_2$, $n_4$, $n_5$, $n_7$, and $n_9$.
Taking $n_5$ and $n_7$, say, as the two independent variables, with $n_7$
constrained by eq. (\ref{ngenhsu10}) to take on values in the set
$\{0,1,...,N_{gh}\}$, we find the formal solution of these equations to be
\beq
n_4 = \frac{1}{3}\left ( 4n_7-N_{gh} \right )
\label{nsol1su10}
\eeq
\beq
n_9 = \frac{4}{3}\left (N_{gh} - n_7 \right ) \ . 
\label{nsol2su10}
\eeq
 From eq. (\ref{nsol2su10}) and the requirement that $n_9$ be a nonnegative
integer, it follows that $N_{gh}-n_7=0$ mod 3.  This could be satisfied for
$N_{gh}=3$ and $n_7=0$, but this choice is excluded because it would
produce a negative value for $n_4$.  The other choice is $N_{gh}=n_7$,
which gives $n_9=0$ and $n_4=n_7$.  Substituting $n_9=0$ in eq. 
(\ref{ngenhsu10}) yields $n_4=N_{gh}$ so that $n_7=N_{gh}$ also;
substituting the latter in eq. (\ref{ngenhsu10}) then gives $n_2=0$.  These
conditions leave $n_5$ free, subject to the additional requirement that
SU(5)$_{SCC}$ be asymptotically free.  Thus, we have
\beq
{\bf n} = (0,0,0,N_{gh},n_5,0,N_{gh},0,0)_{10} \ . 
\label{nsu10}
\eeq
The number of SC-singlet, SU(5)$_{GU}$-singlet fermions is $N_{(1,1)}=2n_5$. 

Let us consider first the value $n_5=0$, so that
\beq
{\bf n} = (0,0,0,N_{gh},0,0,N_{gh},0,0)_{10} \ . 
\label{nsu10h11}
\eeq
This allows one to choose $(N_{gh},N_{g \ell})=(3,0)$, (2,1), or (1,2). 
If one were to apply the irreducibility condition that $GCD(\{ n_k \})=1$, it
would imply that $N_{gh}=1$ in eq. (\ref{nsu10h11}), which 
requires $N_{g \ell}=2$, i.e., the TC group is
SU(3)$_{TC}$ and the SC group should undergo two sequential breakings, ${\rm
SU}(5)_{SC} \to {\rm SU}(4)_{SC}$, followed by ${\rm SU}(4)_{SC} \to {\rm
SU}(2)_{TC}$.  Although we will not impose the irreducibility here, we will
exclude all of the reducible solutions because they lead to excessively many
fermions, which render the SU(8)$_{SCC}$ theory non asymptotically free. 

The fermions in each type of representation have the decomposition, with
respect to the subgroup (\ref{scxgut}) for this case,
\beq
{\rm SU}(5)_{SC} \times {\rm SU}(5)_{GU} \ , 
\label{scxgutsu10}
\eeq
of
\beq
[4]_{10} = (\bar 5,1) + (\overline{10},5) + (10,10) + (5,\overline{10}) 
+ (1,\bar 5)
\label{4_su10_scxgut}
\eeq
\beq
[7]_{10} = [\bar 3]_{10} = (10,1) + (\overline{10},\bar 5) + 
(\bar 5,\overline{10}) + (1,10) \ . 
\label{7_su10_scxgut}
\eeq
Thus, the SU(5)$_{SC}$ sector forms a chiral gauge theory, with left-handed
(nonsinglet) fermion content consisting of
\beq
\{f\} = N_{gh}[10(5 + \overline{10}) + 11(\bar 5 + 10)] \ . 
\label{fsu10ngenh0}
\eeq

With respect to the subgroup (\ref{wxscc}), 
\beq
{\rm SU}(2)_w \times {\rm SU}(8)_{SCC} \ , 
\label{wxsccsu10}
\eeq
the fermions in each representation have the decomposition
\beq
[4]_{10} = (1,[2]_8) + (2,[3]_8) + (1,[4]_8)
\label{4_su10_wcscc}
\eeq
\beq [7]_{10} \approx [\bar 3]_{10} = (1, [\bar 1]_8) + (2,[\bar 2]_8) +
(1,[\bar 3]_8) \ . 
\label{7_su10_wcscc}
\eeq
Some dimensionalities of relevant SU(8) representations are ${\rm
dim}([2]_8)=28$, ${\rm dim}([3]_8)=56$, and ${\rm dim}([4]_8)=70$; note that
$[4]_8=[\bar 4]_8$.  Thus, assuming that the SU(10) unified theory breaks in
such a way as to yield, for a range of lower energies, an effective theory with
SU(8)$_{SCC}$ gauge symmetry, this SCC sector is a chiral gauge theory.  The
SU(8)$_{SCC}$ gauge interaction has leading beta function coefficient
$b^{(SCC)}_0 = 4(22-21N_{gh})/3$, so it is asymptotically free only for the
choice $N_{gh}=1$, for which $b^{(SCC)}_0=4/3$.  As the energy scale decreases
and the coupling $\alpha_{SCC}$ becomes sufficiently large, this theory will
thus form bilinear fermion condensates.  The most attractive channel is
\beq
(1,[4]_8) \times (1,[4]_8) \to (1,1) \ , \quad \Delta C_2 =18 \ . 
\label{su8channel1}
\eeq
This is of the form of channel (\ref{ppchannel}) with $p=4$ and thus violates
weak hypercharge and electric charge.

Other channels with smaller values of $\Delta C_2$ include
\beq
(2,[3]_8) \times (1,[\bar 3]_8) \to (2,1) \ , \quad \Delta C_2 =\frac{135}{8}
\simeq 16.9
\label{su8channel2}
\eeq
\beq
(1,[2]_8) \times (2,[\bar 2]_8) \to (2,1) \ , \quad \Delta C_2 =\frac{63}{4}
= 15.75
\label{su8channel3}
\eeq
\beq
(1,[\bar 3]_8) \times (1,[4]_8) \to (1,[1]_8) \ , \quad \Delta C_2
=\frac{27}{2} = 13.5
\label{su8channel4}
\eeq
\beq
(1,[2]_8) \times (1,[\bar 3]_8) \to (1,[\bar 1]_8) \ , \quad \Delta C_2
=\frac{45}{4} = 11.25 \ . 
\label{su8channel5}
\eeq
Channels (\ref{su8channel2}) and (\ref{su8channel3}) break SU(2)$_w$ at too
high a scale and fail to break SU(8)$_{SCC}$.  Channels (\ref{su8channel4}) and
(\ref{su8channel5}) are allowed by symmetry considerations.  However, since
their $\Delta C_2$ values are smaller than those of the leading channels, one
cannot make a persuasive case that they would occur.  Moreover, because of the
relatively small value of $b^{(SCC)}$, estimates based on eq. (\ref{mucrit})
indicate that for a hypothetical unified coupling $\alpha_{GU} \sim 0.04$ at
$M_{GU} \sim 10^{16}$ GeV, these condensates would form at much too small a
scale for a viable model.

We consider next the choice $N_{gh}=1$, $n_5=1$, so that
\beq
{\bf n} = (0,0,0,1,1,0,1,0,0)_{10} \ . 
\label{n11su10b}
\eeq
For this choice the decomposition of the fermions with respect to the subgroup
(\ref{scxgutsu10}) is 
\beq
[4]_{10} = (\bar 5,1) + (\overline{10},5) + (10,10) + (5,\overline{10}) + 
           (1,\bar 5)
\label{4_su10_scxguta}
\eeq
\beq
[5]_{10} = 2(1,1) + (\bar 5,5) + (5,\bar 5) + (\overline{10},10) + 
(10,\overline{10})
\label{5_su10_scxguta}
\eeq
\beq
[7]_{10} \approx [\bar 3]_{10} = 
(10,1) + (\overline{10},\bar 5) + (\bar 5,\overline{10}) + (1,10) \ . 
\label{7_su10_scxguta}
\eeq
The SU(5)$_{SC}$ theory is a chiral gauge theory, consisting of 
the nonsinglet fermion content
\beq
\{f \} = 16(\bar 5) + 15(5) + 21(10) + 20(\overline{10}) \ . 
\label{f111su10}
\eeq

With respect to the subgroup (\ref{wxsccsu10}) the fermions in the 
$[4]_{10}$ and $[7]_{10}$ have the decompositions given in eqs. 
(\ref{4_su10_wcscc}) and (\ref{7_su10_wcscc}), and 
\beq
[5]_{10} = (1,[3]_8) + (2,[4]_8) + (1,[\bar 3]_8) \ . 
\label{5_su10_wxscca}
\eeq
Although the $[5]_{10}$ is self-conjugate, the $[4]_{10}$ and $[7]_{10}$ make
this SU(8)$_{SCC}$ theory chiral.  However, it is non-asymptotically
free, with leading beta function coefficient $b^{(SCC)}_0=-22$.

Finally, we consider the choice $N_{gh}=0$, $n_5=1$, so that
\beq
{\bf n} = (0,0,0,0,1,0,0,0,0)_{10} \ . 
\label{n11su10c}
\eeq
For this choice the decomposition of the fermions in the $[5]_{10}$ with
respect to the subgroup (\ref{scxgutsu10}) is given by
eq. (\ref{5_su10_scxguta}) and with respect to the subgroup (\ref{wxsccsu10})
by eq. (\ref{5_su10_wxscca}).  Assuming that the breaking of SU(10) is such as
to yield an SU(8)$_{SCC}$-invariant theory for a range of lower energies, its
coupling does grow as the energy decreases, as governed by the leading beta
function coefficient $b^{(SCC)}_0 = 6$.  However, as is evident from
eq. (\ref{5_su10_wxscca}), this theory is vectorial, so that when the
SU(8)$_{SCC}$ coupling grows sufficiently large to produce a fermion
condensate, this condensate will preferentially preserve the SU(8)$_{SCC}$
symmetry rather than breaking it, as is necessary, to ${\rm SU}(5)_{TC} \times
{\rm SU}(3)_c$ (and thence, sequentially, breaking SU(5)$_{TC}$ to
SU(2)$_{TC}$).  We have investigated higher values of $N_{SC}$ and thus $N$ but
have found that they exhibit problems similar to those of the models above.

\subsection{Assessment} 

Thus we we find several general problems with the unification approach embodied
in models of type 3. One, pertaining to the mechanism for breaking the unified
gauge group $G$ in a weak-coupling framework, also applies to models of type
(1) and (2) and will be discussed in the next section.  A second problem is
that, even if one could arrange some mechanism to break the $G$ symmetry in the
desired manner to yield a lower-energy theory invariant, presumably, under an
SCC gauge symmetry combining the strongly coupled group $G_{SC}$ with the color
group, it appears very difficult to get this SCC symmetry to break in the
requisite manner.  This is especially true when, as is often the case, the
SU($N_{SCC}$) gauge interaction is vectorial.  Even when it is chiral (and
asymptotically free), as in the model with $N_{SC}=5$ and ${\bf n}$ given by
eq. (\ref{nsu10h11}) with $N_{gh}=1$, the most attractive condensation channels
do not lead to the requisite breaking.

Since the determination of the resultant SC and TC sectors depends on having a
viable SCC breaking pattern, this prevents one from proceeding very far with
the analysis of these lower-energy effective field theories in the context of
these models. However, we note that in cases where $N_{g \ell} \ge 1$, it could
also be challenging to get the SC symmetry to break sequentially down to the
resultant exact TC symmetry.  Because of the previous problems, one cannot
obtain very definite predictions for fermion masses.  A general concern
pertains to reproducing the observed mass hierarchy of the three SM fermion
generations.  In order to do this, one tends to need three different ETC mass
scales, essentially the $\Lambda_j$, $j=1,2,3$ discussed in Section II.  In the
present approach, one has $N_{g \ell}+1$ ETC-type mass scales.  This is
illustrated by the scenario in which the SU($N_{SCC}$) symmetry first breaks to
${\rm SU}(N_{SC}) \times {\rm SU}(3)_c$ and then ${\rm SU}(N_{SC})$ breaks
sequentially at $N_{g \ell}$ lower scales, finally yielding the residual exact
SU($N_{TC}$) symmetry.  Consequently, unless $N_{g \ell} \ge 2$, one does not
have enough mass scales to account for the SM fermion mass hierarchy.  This
problem reaches its most acute form when $N_{g \ell}=0$, $N_{gh}=3$.

Because of the presence of intermediate scales between $m_Z$ and $M_{GU}$ with
nonperturbative behavior and the feature that the ETC gauge bosons involved in
symmetry breakings at these scales carry SM quantum numbers, the calculation of
the evolution of the SM gauge couplings is more complicated in models of type
(3) than in models of types (1) or (2).  However, exploratory analysis of
plausible evolution of gauge couplings for the various scenarios that we have
examined indicate that satisfying the constraint of gauge coupling unification
is still a very restrictive requirement.

\section{Dynamical Breaking of Unified Gauge Symmetries}
\label{section6} 

In addition to other issues that we have addressed concerning prospects for
unification of gauge symmetries in a dynamical context, there is another one
which is quite general.  For the present discussion let us assume that one has
a model that does achieve gauge coupling unification.  The resultant value of
the unified gauge coupling at $M_{GU}$ is generically expected to be small.
But if one is trying to construct a theory in which all gauge symmetry breaking
is dynamical, this would normally require there to be some strongly coupled
gauge interaction at the relevant scale.  For example, the dynamical breaking
of the electroweak symmetry in a technicolor theory requires that technicolor
be an asymptotically free gauge interaction that becomes strongly coupled at
the electroweak scale.  Although the ETC interaction is strongly coupled at the
scale $\Lambda_1 = \Lambda_{ETC, max} \simeq 10^6$ GeV, in typical models of
type (1) and (2) the ETC coupling evolves to relatively small values as the
energy scale ascends to the region of $M_{GU}$, so one could not use ETC
interactions to break $G_{GU}$ at $M_{GU}$ in these models.  Moreover, in
specific models such as those of Refs. \cite{at94}-\cite{kt}, the ETC (and HC)
condensates involve SM-singlet fields which, in the present context, would
naturally be $G_{GU}$-singlet fields, so their condensates would not break
$G_{GU}$.

One way to break a hypothetical symmetry $G_{GU}$ unifying SM gauge
interactions at a high scale $M_{GU}$ would be to expand the theory to include
an additional gauge interaction, associated with a group $G_a$, that is
strongly coupled at this scale, together with fermions that transform as
nonsinglets under both $G_{GU}$ and $G_a$.  The great disparity between the
coupling strengths of the $G_{GU}$ and $G_{ETC}$ gauge bosons, on the one hand,
and the assumed $G_a$ gauge bosons, on the other, is a striking property of
such a model.  Another approach would be to envision a nonperturbative
unification of gauge symmetries, but by its very nature, this is difficult to
study reliably using tools such as perturbative evolution of SM couplings
\cite{nonpert}.  

For completeness, one should note that a major objection to the use of Higgs
fields for symmetry breaking at lower scales is the instability of a Higgs
sector to large radiative corrections, necessitating fine tuning to keep the
Higgs bosons light compared with an ultraviolet cutoff.  But from
considerations of nucleon stability alone, not to mention gauge coupling
evolution, one knows that in the (four-dimensional) theories considered here,
$M_{GU}$ is generically quite high, not too far below the Planck scale where
the theories are certainly incomplete, since they do not include quantum
gravity.  Hence, this objection would not be as strong for the breaking of
$G_{GU}$ as for symmetry breakings that occur at substantially lower scales.

\section{Conclusions}
\label{conclusions} 

In this paper we have analyzed approaches to the partial or complete
unification of gauge symmetries in theories with dynamical symmetry breaking.
We considered three main types of models with progressively greater degrees of
unification, including those that (1) involve sufficient unification to
quantize electric charge, (2) attempt to unify the three standard-model gauge
interactions in a simple group that forms a direct product with an extended
technicolor group, and, (3) attempt to unify the standard-model gauge
interactions with (extended) technicolor in a simple group.  The model of
Refs. \cite{lrs,nag} is a successful example of theories of type (1).  Models
of type (3) provide an interesting contrast to those of types (1) and (2) in
the different way in which standard-fermion generations are produced and in the
property of having ETC gauge bosons that carry standard-model quantum numbers.
We have pointed out a number of challenges that one faces in trying to
construct viable models of types (2) and (3).  There are certainly further
avenues for research in this area.  For example, one could investigate direct
product groups involving auxiliary hypercolor-type groups. Another idea would
be to study ways to unify top-color \cite{tc2} with standard-model gauge
symmetries. In conclusion, it is possible that electroweak symmetry breaking is
dynamical, involving a new strongly coupled gauge symmetry, technicolor, 
embedded in an extended technicolor theory to give masses to standard-model 
fermions.  There is a strong motivation to understand how the associated
symmetries can be unified with the color and weak isospin and hypercharge gauge
symmetries.  We hope that the results of the present paper will be of use in
the further study of this unification program. 

\bigskip
\bigskip

We thank T. Appelquist for helpful discussions, in particular for the
collaborative work on the partial unification model in Refs. \cite{lrs,nag}.
This present research was partially supported by the grant NSF-PHY-00-98527.

\bigskip
\bigskip

\section{Appendix}

We gather in this appendix some standard formulas that are used in the
calculations reported in the text.

\subsection{Some Group-Theoretic Properties of Representations $[k]_N$ of
SU($N$) }

We denote the completely antisymmetric $k$-fold products of the fundamental and
conjugate fundamental representation of SU($N$) as $[k]_N$ and $[\bar k]_N$,
respectively.  These can be displayed as tensors with $k$ upper indices,
$\psi^{i_1...i_k}$ and $k$ lower indices, $\psi_{i_1...i_k}$, and have the
(same) dimension
\beq
{\rm dim}([k]_N) = {N \choose k} \ . 
\label{dimkn}
\eeq
These representations satisfy the equivalence property
\beq
[N-k]_N \approx [\bar k]_N
\label{nkn}
\eeq
under SU($N$), as follows by contraction with the totally antisymmetric tensor
density $\epsilon_{i_1...i_N}$.  The fact that these representations have the
same dimension is evident from the identity ${N \choose k} = {N \choose N-k}$.
Further, if $N$ is even, say $N=2k$, then $[N/2]_N$ is 
self-conjugate with respect to SU($N$). 

The contribution of the (left-handed) fermions in the representation $[k]_N$ to
the gauge anomaly for SU($N$) is \cite{g79,anomaly}
\beq
A([k]_N)=\frac{(N-2k)(N-3)!}{(N-k-1)!(k-1)!} \ , 
\label{anomaly}
\eeq
where the normalization is such that contribution of the fundamental
representation is 1.  The property $A([k]_N)=-A([\bar
k]_N)=-A([N-k]_N)$ for $1 \le k \le N-1$ is evident in
eq. (\ref{anomaly}).

The quadratic Casimir invariant $C_2({\cal R})$ for the representation
${\cal R}$ is defined by 
\beq
\sum_{a=1}^{o(G)} \sum_{j=1}^{{\rm dim}({\cal R})}
  {\cal D}_{\cal R}(T_a)^i_j{\cal D}_{\cal R}(T_a)^j_k = C_2({\cal
R})\delta^i_k \ , 
\label{c2r}
\eeq
where ${\cal D}_{\cal R}(T_a)$ is the ${\cal R}$-representation of the
generator $T_a$ and $o(G)$ is the order of the group $G$.

The contribution of a fermion loop, for fermions of representation ${\cal R}$
of SU($N$), to the beta function coefficient $b^{(j)}_0$, involves the
invariant $T({\cal R})$ defined by
\beq
\sum_{i,j=1}^{{\rm dim}({\cal R})} 
({\cal D}_{\cal R}(T_a))^i_j \ ({\cal D}_{\cal R}(T_b))^j_i = T({\cal R}) 
\delta_{ab} \ . 
\label{tr}
\eeq
These invariants satisfy the elementary relation 
\beq
C_2({\cal R}) \ {\rm dim}({\cal R}) = T({\cal R}) \ o(G) \ . 
\label{c2tr}
\eeq
For SU($N$), 
\beq
C_2([k]_N) = \frac{k(N+1)(N-k)}{2N} 
\label{c2kn}
\eeq
and 
\beq
T([k]_N) = \frac{1}{2}{N-2 \choose k-1} \ . 
\label{tkn}
\eeq

As the value of $N$ (in eq. (\ref{n})) increases, the number of (left-handed)
fermions in a generic set ${\bf n}$ tends to increase rapidly.  For example, 
for even $N_{SC}$ and hence odd $N=N_{SC}+5 \equiv 2m+1$, the set ${\bf
n}$ with $n_{2\ell} = 1$ for $\ell=1,..m$ and
$n_{2\ell+1}=0$ for $\ell=0,..,m-1$, has a total number of fermions given by 
\beq
\sum_{\ell=1}^m {2m+1 \choose 2\ell} = 2^{2m}-1 = 2^{N-1}-1 \ . 
\label{nftot}
\eeq
Evidently, this grows exponentially rapidly with $N_{SC}$ and $N$, which
quickly renders the SCC theory non-asymptotically free.  Note that for this
choice of ${\bf n}$, since $n_2=n_{_{N_{SC}+2}}$, it follows that $N_{gh}=2$.

\subsection{Formulas for the Evolution of Gauge Couplings} 

We consider a factor group $G_j$ with gauge coupling $g_j$ and denote $\alpha_j
=g_j^2/(4\pi)$.  The evolution of the gauge couplings as a function of the
momentum scale $\mu$ is given by the renormalization group equation
\beq
\beta_j = \frac{d \alpha_j}{dt} = - \frac{\alpha_j^2}{2\pi}\left ( b_0^{(j)} +
\frac{b_1^{(j)}}{2\pi}\alpha_j + O(\alpha_j^2) \right ) \ , 
\label{beta}
\eeq
where $t=\ln \mu$, and the first two terms $b_0^{(j)}$ and $b_1^{(j)}$ are
scheme-independent. The beta function with perturbatively calculated
coefficients is appropriate to describe the running of the respective couplings
in the energy ranges where the respective gauge fields are dynamical (i.e.,
above corresponding scales at which $G_j$ is broken) and where the couplings
$\alpha_j$ are not too large.  For our analyses of the perturbative evolution
of gauge couplings, it will be sufficient to keep only the $b_0^{(j)}$ term;
the well-known solution of eq. (\ref{beta}) is then given by
\beq
\alpha_j^{-1}(t_2) = \alpha_j^{-1}(t_1) + \frac{b^{(j)}_0}{2\pi}(t_2 - t_1) \ .
\label{alfsol}
\eeq
If an effective field theory involves the direct product of two gauge 
groups $G_j$ and $G_k$ for energy scales between $\mu_\ell$ and a larger
scale $\mu_{jk}$ where the associated couplings $\alpha_j$ and $\alpha_k$ are
equal, one has, to this order,
\beq
\ln \left ( \frac{\mu_{jk}}{\mu_\ell} \right ) = 
\frac{2\pi[\alpha_j^{-1}(\mu_\ell) - \alpha_k^{-1}(\mu_\ell)]}
     {b^{(k)}_0 - b^{(j)}_0} \ . 
\label{mujk}
\eeq
The three SM gauge couplings are accurately determined at $\mu = m_Z$, with the
results \cite{pdg,ew} $\alpha_3(m_Z) \simeq 0.118$, $\alpha_{em}(m_Z)^{-1}
\simeq 128$, and $(\sin^2\theta_W)_{\overline{MS}}(m_Z) \simeq 0.2312$. The
SU(2)$_L$ and U(1)$_Y$ couplings $g \equiv g_{2L}$ and $g' \equiv g_Y$ are
given by $e = g \sin \theta_W = g' \cos \theta_W$ and have the values (quoted
to sufficient accuracy for our present purposes) $\alpha_2(m_Z) = 0.033$ and
$\alpha_Y(m_Z) = 0.010$.  The evolution of these couplings to scales $\mu >
m_Z$ depends on the type of gauge symmetry unification that one is considering.

\subsection{Fermion Condensation}

Consider massless fermions that transform according to some representations
${\cal R}$ of a nonabelian gauge group $G$. In the approximation of
a single-gauge boson exchange , the critical coupling for condensation in the
channel 
\beq
{\cal R}_1 \times {\cal R}_2 \to R_{cond.}
\label{channel}
\eeq
is given by \cite{gap} 
\beq
\alpha_c = \frac{2\pi}{3 \Delta C_2} 
\label{alfcrit}
\eeq
where
\beq
\Delta C_2 = C_2({\cal R}_1) + C_2({\cal R}_2)-C_2({\cal R}_{cond.}) \ . 
\label{deltac2}
\eeq
Because $\alpha \sim O(1)$ where fermion condensation occurs, the one-gauge
boson approximation is only a rough guide to the actual critical value of
$\alpha$.  Corrections to this have been estimated in Ref. \cite{gap}.  In
addition to gauge boson exchange diagrams, nonperturbative processes involving
instantons are also important \cite{instanton}.

For condensation due to the asymptotically free gauge interaction with gauge
group $G_j$ and associated squared coupling $\alpha_j$ obeying a
renormalization group equation with leading beta function coefficient
$b_0^{(j)}$, the solution in eq. (\ref{alfsol}) together with the condition
(\ref{alfcrit}), yield, for the mass scale at which the condensation takes
place, the rough estimate 
\beq
\mu_{c,j} \simeq M_{GU} \ {\rm exp} \left [ -\frac{2\pi}{b_0^{(j)}} 
\left ( \alpha_j(M_{GU})^{-1} - \frac{3\Delta C_2}{2\pi} \right ) \right ]
\label{mucrit} 
\eeq
where $\Delta C_2$ is the value appropriate for this channel, as given by
eq. (\ref{deltac2}).

\vspace{300cm}

\newpage

\begin{table}
\caption{\footnotesize{Some properties of the various models of type (3)
discussed in the text with $G_{SC}$ and $G_{SM}$ unified in a simple group $G$.
Here, $G_{SC}={\rm SU}(N_{SC})$, $G_{TC}={\rm SU}(N_{TC})$, and $G_{SC}
\supseteq G_{TC}$.  The column marked ``SCC'' lists some properties of the
SU($N_{SCC}$) theory combining the ${\rm SU}(N_{SC})$ and SU(3)$_c$ groups.
See text for further definitions and discussion.  The fermion content is
indicated by the vector ${\bf n}$ (with subscript omitted for brevity).  The
notation ``no sol.'' means that (in the dynamical framework used) there is no
solution to the requirements of anomaly freedom, well-defined SM fermion
generations, and $N_{gen.}=3$. The notation VGT and $CGT$ indicate that the
gauge interaction is vectorial and chiral, respectively; AF and NAF mean
asymptotically free and non asymptotically free, respectively. The number
$N_{(1,1)}$ in column 9, given by eq. (\ref{n11}), is the number of
electroweak-singlet neutrinos.}}
\begin{center}
\begin{tabular}{|c|c|c|c|c|c|c|c|c|} \hline\hline
$N$ & $N_{SC}$ & $N_{SCC}$ & $N_{TC}$ & $N_{g\ell}$ & $N_{gh}$ & 
${\bf n}$ & SCC & $N_{(1,1)}$ \\ \hline
7  & 2 & 5 & 2 & 0 & 3 & no sol.     & $-$        & $-$   \\ \hline
8  & 3 & 6 & 3 & 0 & 3 & (0200103)   & VGT, \ AF  & 1     \\ \hline
8  & 3 & 6 & 2 & 1 & 2 & no sol.     & $-$        & $-$   \\ \hline
9  & 4 & 7 & 4 & 0 & 3 & no sol.     & $-$        & $-$   \\ \hline
9  & 4 & 7 & 3 & 1 & 2 & (01010101)  & VGT, \ AF  & 1     \\ \hline
9  & 4 & 7 & 2 & 2 & 1 & no sol.     & $-$        & $-$   \\ \hline   
10 & 5 & 8 & 5 & 0 & 3 & (000300300) & CGT, \ NAF & 0     \\ \hline   
10 & 5 & 8 & 4 & 1 & 2 & (000200200) & CGT, \ NAF & 0     \\ \hline   
10 & 5 & 8 & 3 & 2 & 1 & (000100100) & CGT, \ AF  & 0     \\ \hline
10 & 5 & 8 & 3 & 2 & 1 & (000110100) & CGT, \ NAF & 2     \\ \hline
10 & 5 & 8 & 2 & 3 & 0 & (000010000) & VGT, \ AF  & 2    \\ \hline\hline
\end{tabular}
\end{center}
\label{properties}
\end{table}


\begin{thebibliography}{99}
\bibitem{simple} 
Henceforth in the paper we shall omit the technical distinction between simple
and semisimple groups. Recall that a group is (semi)simple if it contains no
(abelian) invariant subgroup other than the identity.  For the present study
the important property of a gauge theory involving a (semi)simple gauge group
is that it only involves a single gauge coupling.
\bibitem{b0cases} 
The values of the leading beta function coefficient
$b^{(ETC)}_0$ (as defined in eq. (\ref{beta})) for the ETC gauge coupling
depend on SM-singlet, as well as SM-nonsinglet, ETC-nonsinglet fermions.  For
the recent SU(5)$_{ETC}$ models, the ETC-nonsinglet, SM-singlet sectors
consisted of the following fermion representations with respect to the group
${\rm SU}(5)_{ETC} \times {\rm SU}(2)_{HC}$: (a) $(10,1)_L$ and
$(\overline{10},2)_L$ in Ref. \cite{nt}; (b) $(\overline{10},1)_L$ and
$(\overline{10},2)_L$ in Refs. \cite{ckm,kt}; and (c) the ETC-nonsinglet,
$G_{422}$-singlet fermions $(5,1)_L$, $(10,1)_L$, and $(\overline{10},2)_L$ in
Refs. \cite{lrs,nag}. The resulting values of the leading coefficient
$b^{(ETC)}_0$ of the beta function for SU(5)$_{ETC}$ are, respectively, (a)
31/3, (b) 31/3, and (c) 29/3.
\bibitem{tcvac}
This vacuum alignment argument is used, for example, to infer that, although
all of the bilinear technifermion condensates $\langle \bar F F \rangle$ (and, 
for SU(2)$_{TC}$, also $\langle \epsilon_{jk} F^{j \ T} C F^k \rangle$) are
equivalent from the point of view of technicolor alone, the ones that actually
form are only those that conserve color and electric charge, and not
condensates such as $\langle \bar U D \rangle$, $\langle \bar U E \rangle$,
etc. that violate these symmetries.
\bibitem{bcond} 
As will be discussed later in the text, the requirement in this
old approach that $\beta_{TC} < \beta_{SU(3)_c} < 0$ is replaced in our
analysis by the requirement of asymptotic freedom of the SCC sector unifying 
the SC and color interaction.
\bibitem{nonpert} 
Related possibilities arise in models with large extra dimensions, where gauge
couplings have power-law, rather than logarithmic, dependence on scale; we
do not consider these here.
\end{thebibliography}
\end{document}